\documentclass[aps,prl,twocolumn,secnumarabic,superscriptaddress,longbibliography]{revtex4-2}

\usepackage{epsf}
\usepackage{bm}
\usepackage{hyperref}
\usepackage{amsfonts}
\usepackage{amssymb}
\usepackage{amsmath,mathtools}
\usepackage{array}
\usepackage{enumerate,dsfont}
\usepackage{dcolumn,multirow}
\usepackage[utf8]{inputenc}
\usepackage{latexsym}
\usepackage{xcolor}
\usepackage{tikz-cd}
\usepackage{ulem}
\renewcommand{\emph}[1]{{\it #1}}

\usepackage{lipsum}

\begin{document}

\title{Local-to-Global Entanglement Dynamics by Periodically Driving Impurities}

\author{Zhi-Xing Lin}
\affiliation{Department of Physics, Princeton University, Princeton, New Jersey, 08544, USA}

\author{Abhinav Prem}
\affiliation{Physics Program, Bard College, 30 Campus Road, Annandale-on-Hudson, New York 12504, USA}
\affiliation{School of Natural Sciences, Institute for Advanced Study, Princeton, New Jersey 08540, USA}

\author{Shinsei Ryu}
\affiliation{Department of Physics, Princeton University, Princeton, New Jersey, 08544, USA}

\author{Bastien Lapierre}
\affiliation{Department of Physics, Princeton University, Princeton, New Jersey, 08544, USA}
\affiliation{Philippe Meyer Institute, Physics Department, École Normale Supérieure (ENS), Université PSL, 24 rue Lhomond, F-75231 Paris, France}

\date{June 19, 2026}

\begin{abstract}
We study the entanglement dynamics of one-dimensional fermionic chains subject to a local Floquet drive of a two-site impurity, and uncover a sharp transition in the entanglement dynamics set by the driving period $T$. For large periods, the entanglement entropy (EE) grows linearly in time, signaling a heating phase with volume-law entanglement; below a critical period $T_\ast$, the EE instead grows subextensively, characteristic of a local quantum quench. We establish this dichotomy in two complementary settings: a gapless nearest-neighbor hopping chain, where a single transition separates logarithmic from volume-law growth, and a gapped Su–Schrieffer–Heeger (SSH) chain, whose two-band structure yields a richer phase diagram with multiple area-to-volume-law transitions. In the noninteracting limit, we trace these transitions analytically to quasienergy folding in the single-particle Floquet spectrum: a single $\pi$-gap closure for the NN chain, and a sequence of foldings at both $0$- and $\pi$-gaps for the SSH chain, yielding the alternating pattern of heating and non-heating phases. We further show that the so-called ``average energy" operator furnishes a many-body diagnostic of the transition, remaining local in the non-heating phase but developing non-local couplings in the heating phase. For the gapless chain, using extensive matrix-product-state simulations, we demonstrate that the non-heating phase and its subextensive entanglement growth survive weak interactions over numerically accessible timescales. Our results establish local Floquet engineering as a route to emergent bulk phenomena, offering a new perspective on energy localization and thermalization in driven many-body systems.

\end{abstract}
\maketitle


\textit{Introduction} --- Local impurities and defects can play a central role in shaping the global properties of quantum many-body systems. In one spatial dimension (1d), even a single localized scatterer can dramatically alter low-energy physics, producing phenomena such as orthogonality catastrophes~\cite{anderson1967} or impurity-driven quantum phase transitions~\cite{vojta2006impurity,bayat2017scaling,rancati2020,huang2020quantum}. 
For 1d free-fermionic systems, the impact of localized impurities on many-body correlations, transport, and entanglement dynamics is well-studied both in equilibrium~\cite{Peschel2005,zhao2006,lin2009ent,Sakai_2008,klich2009quantum,calabrese2011ent,calabrese2012ent,Peschel_2012,gupterle2017,kruthoff2021,rogerson2022,roy2022ent} and in the context of out-of-equilibrium quench dynamics~\cite{eisler2007,song2011ent,Eisler_2012,Wen_2018_interface,Gruber_2020,PhysRevB.103.L041405,Capizzi_2023,Capizzi2023ent,saha2024,Collura_2013,vasseur2013,kennes2014uni,mitra2015transport, prosen2019non,gamayun2020,deluca2020,mitraquenchreview}. Such studies have revealed e.g., that the entanglement entropy (EE) in critical spin chains grows linearly in time after a global quench across a conformal interface~\cite{Wen_2018_interface,Capizzi2023ent}.

\begin{figure}[t]
	\includegraphics[width=0.4\textwidth]{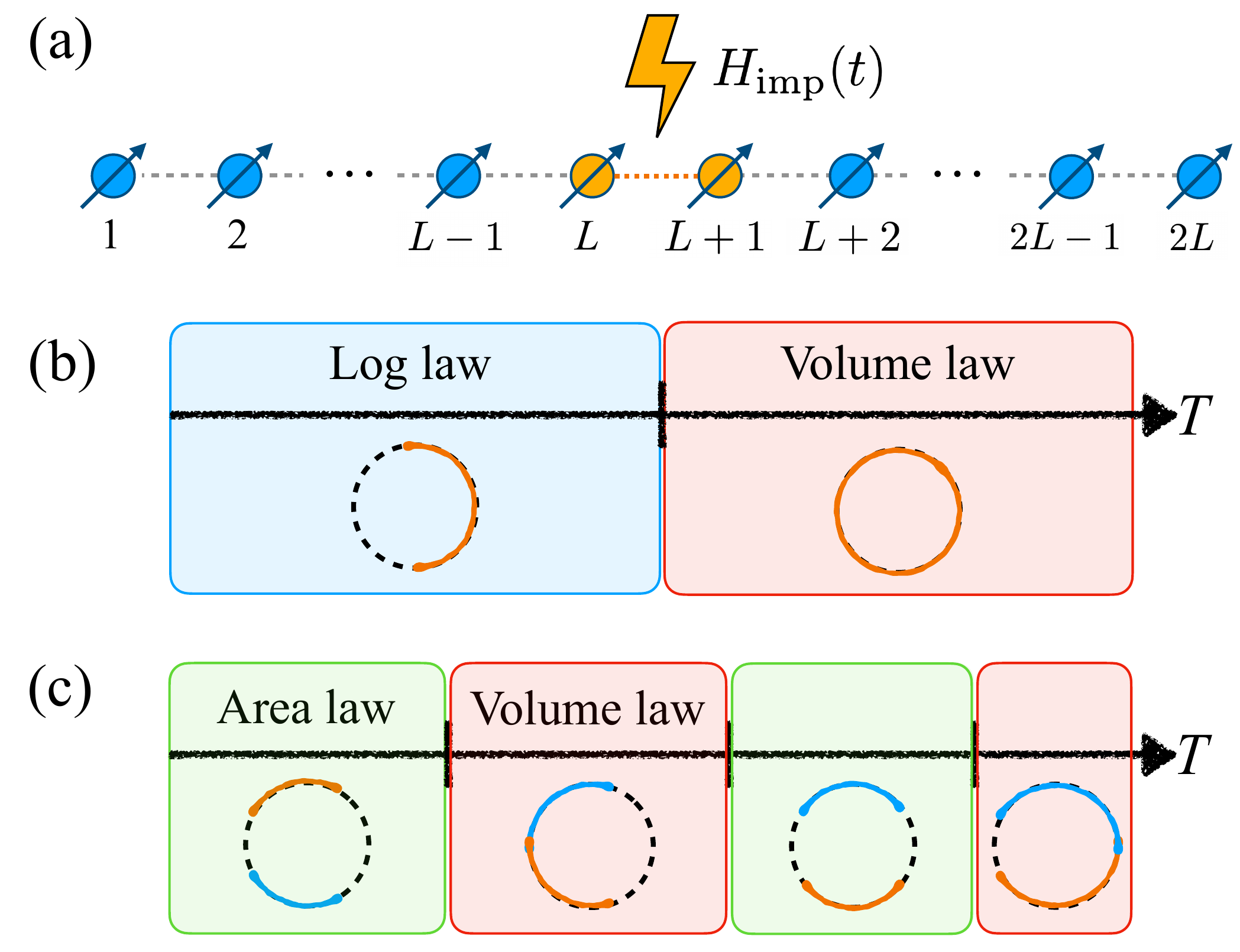}
	\caption{\label{fig1_impurity} (a) Setup: We consider a periodically driven two-site impurity (in yellow) immersed in a uniform chain of length $2L$. (b-c) Main result: Below a critical period $T_*$, the drive only acts locally and leads to local quench dynamics characterized by a subextensive growth of EE (`local quench' scenario). Above this critical period, the drive acts as a non-local perturbation which leads to linear growth of EE (`global quench' scenario). In the noninteracting setting, the entanglement transition corresponds to a gap closure of the quasienergy spectrum. (b) In the gapless scenario, a single bulk gap closure happens, separating logarithmic and volume-law entanglement growth.  (c) In the gapped scenario, multiple transitions arise due to quasienergy foldings between bulk bands, separating area and volume-law phases.}
\end{figure}

Periodic (or Floquet) driving presents a qualitatively different regime by introducing another scale into the system (the driving period) and opens the door to distinct dynamical phenomena~\cite{oka2019review,khemani2019review,else2020review,harper2020review,rudner2020band}. Although Floquet dynamics can host novel many-body phases, they also generically induce heating in clean, interacting systems~\cite{D_Alessio_2014,Lazarides2014,ikeda2021fermi,rakcheev2022Floquet}, thereby typically inducing a featureless infinite temperature state. Much of the existing theory has focused on global Floquet drives, where the entire system is periodically driven, leading to distinct heating and non-heating phases and, in conformal limits, to Floquet conformal field theories~\cite{Prosen_2011,russomanno2012,D_Alessio_2013, Citro_2015,bukov2016,Gritsev_2017,PhysRevLett.118.260602,wen2018fcft,fan2020, PhysRevResearch.2.023085, PhysRevB.103.224303,haldar2021,bukov2021pre,gritsev2023,PhysRevB.103.L041405,ikeda2024,Bhattacharjee_2024,gadge2025temp,lin2025chiral}. Recently, it has been appreciated that even \textit{strictly local} Floquet drives (which explicitly break translation symmetry) can dramatically alter global properties, such as transport or thermalization time-scales~\cite{thuberg2016,reyes2017,agarwala2017,PhysRevB.103.L041405,hubner2022,hubner2023,prem2023,melendrez2024,mukherjee2024emergent,li2025dynamics}, thereby opening new avenues for reshaping the bulk dynamical phases of extended systems via local Floquet driving.

In this Letter, we investigate the fate of entanglement dynamics under a locally-driven impurity in an otherwise uniform 1d tight-binding fermion chain. By exploring both a gapless nearest-neighboring (NN) hopping chain and a gapped Su–Schrieffer–Heeger (SSH) chain, we uncover that periodically driving a local defect produces a sharp entanglement transition as the Floquet period $T$ is varied (see Fig.~\ref{fig1_impurity}). 
Below a critical period $T_*$, the driven impurity behaves as a local quench, resulting in subextensive growth of the EE in time; however, for $T > T_*$, the drive effectively couples spatially distant parts of the chain and entanglement grows linearly, characteristic of global Floquet heating. While the gapless NN chain features a single transition, the gapped SSH chain exhibits a richer phase diagram with multiple transition points. Hence, we show that even a local Floquet drive can trigger heating and volume-law entanglement across the entire system.

We trace the origin of this heating transition to a spectral transition in the single-particle Floquet unitary $U_\text{F}$ and also in the many-body ``average energy'' operator.
In the noninteracting limit, $U_\text{F}$ exhibits a $\pi$-gap closure precisely at $T=T_\ast$: while the dynamics remain effectively local below $T_\ast$, the quasienergy folding beyond $T_\ast$ introduces non-local resonance between distinct eigenmodes, driving rapid entanglement growth.
Adopting a recent geometric formulation of Floquet theory~\cite{schindler2025geometricfloquettheory}, we provide a many-body diagnostic of this transition by interpreting it as an energy localization transition~\cite{D_Alessio_2013}. Specifically, in the non-heating phase ($T<T_*$), the initial state remains localized around a particular Floquet eigenstate--the ground-state of the average energy operator--while above $T_*$, the drive begins to populate a broad range of Floquet eigenstates, causing rapid heating.
Finally, we employ extensive matrix-product-state (MPS) simulations to show that these features survive up to numerically accessible timescales, when weak interactions are added to the gapless NN model. We emphasize that, while reminiscent of entanglement transitions in monitored circuits, the phenomenon observed here is distinct and arises from purely unitary dynamics.


\textit{Driven impurity} --- 
We consider a 1d free-fermion chain of length $2L$ with a periodically driven local impurity at the center, as illustrated in Fig.~\ref{fig1_impurity}(a). The bulk chain is static, whereas the impurity part is modulated periodically, such that $H(t+T)=H(t)$. Two representative bulk Hamiltonians are investigated: a gapless NN hopping chain and the gapped SSH model. We initialize the system from the ground state of the impurity-free Hamiltonian at half filling, whose EE follows a logarithmic scaling for the gapless model and an area law for the gapped SSH chain. 
Our goal is to provide a physical picture for the EE growth $S_A(t)=-\text{Tr}(\rho_A\log \rho_A)$. The Hamiltonian reads,
\begin{equation}
\begin{aligned}
\label{eq:floquetdrive_free}
H(t) =& - \frac{1}{2} \left( \sum_{j=1}^{L-1} +\sum_{j=L+1}^{2L-1} \right)\left(t_j c_{j+1}^{\dagger}c_j+h.c.\right) \\
&+ \left(c^{\dagger}_L,\ c^{\dagger}_{L+1} \right)\Lambda(t)\begin{pmatrix} c_L \\ c_{L+1} \end{pmatrix} \, .
\end{aligned}
\end{equation}
Here,
$c_j,c_j^\dagger$ are spinless fermion annihilation/creation operators (with $\{c_j, c_k^\dagger\}=\delta_{jk}$), and $t_j$ denotes the hopping amplitudes: (i) $t_j= 1 $ for the gapless NN model, and (ii) $t_j = 1 - (-1)^j \delta $ for the gapped SSH model with $\delta$ the dimerization parameter.
The matrix $\Lambda(t)=\Lambda(t+T)$ encodes the time-dependence of the impurity. A special case is the conformal defect~\cite{Sakai_2008, Peschel_2012, Eisler_2012, Wen_2018_interface, Capizzi2023ent, PhysRevB.103.L041405, Capizzi_2023, Gruber_2020,saha2024}
\begin{equation}
    \Lambda(t)=\frac{t_L}{2}\begin{pmatrix}
\sqrt{1-\lambda^2(t)}& -\lambda(t)\\ -\lambda(t)& -\sqrt{1-\lambda^2(t)}
\end{pmatrix},
\end{equation}
with $\lambda(t)\in [-1,1]$ such that $\lambda(t+T)=\lambda(t)$. This Hamiltonian interpolates between the defect-free chain and two decoupled chains of length $L$. This form of defect is called conformal because for static $\lambda(t)=\lambda$ and uniform hopping ($t_j = 1$), Eq.~\eqref{eq:floquetdrive_free} is well described by an interface CFT in the low-energy regime~\cite{Sakai_2008} and thus provides a simple model for a gapless theory with a defect. 
Crucially, the findings reported here do not depend on the precise form of $\Lambda(t)$, and even hold for non-Hermitian impurities (see Supplemental Material (SM)~\cite{supmat}).

While there has been significant interest in understanding the dynamics of entanglement after a \textit{local} quench induced by such a defect~\cite{Collura_2013, Capizzi2023ent}, the effect of time-dependent defects on entanglement dynamics has thus far not been studied. As one might anticipate, a sufficiently low frequency periodic drive will couple modes non-locally throughout the whole chain, typically resulting in a linear growth of EE. However, as we will show, there exists a \textit{finite} periodicity threshold $T_*$, set by the single-particle bandwidth, below which the effect of the driven impurity \textit{remains local}. The resulting dynamics is similar to that induced by a local quantum quench~\cite{Calabrese_2007}, and thus only produces subextensive entanglement growth. 
In the following, we provide different viewpoints on the resulting entanglement transition, and discuss its stability under weak interactions.

\begin{figure}
    \centering
    \includegraphics[width=\linewidth]{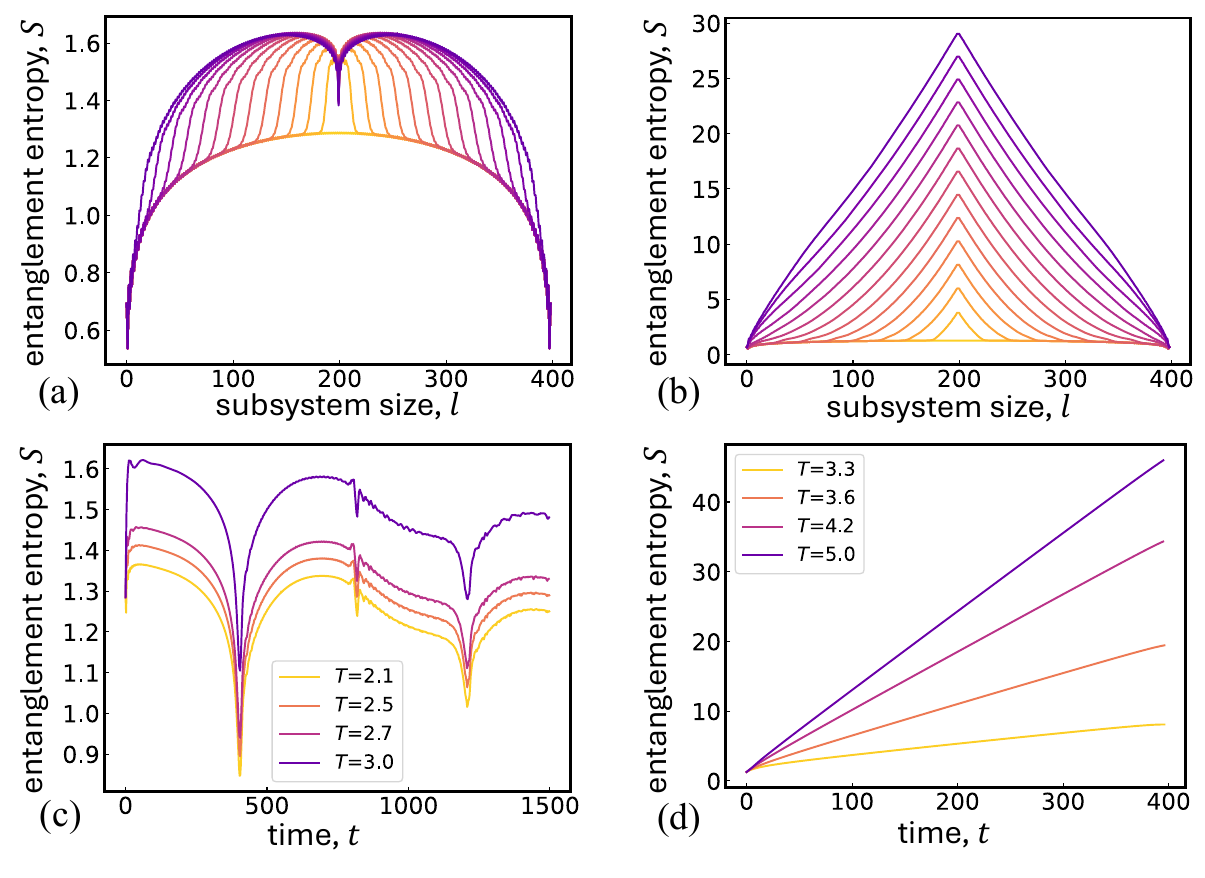}
    \caption{The entanglement phases of the gapless NN model ($2L=400$) subject to a 2-step driven impurity (Eq.~\eqref{eq:floquetdrive_free}) with $\lambda_0=0.5$. In the non-heating phase ($T<\pi$), (a) the EE exhibits logarithmic scaling over time (plotted every six Floquet cycles at $T=2.5$, from orange to purple) and (c) the half-partition EE shows partial revivals that are independent of the driving frequency. In contrast, in the heating phase $(T > \pi)$, (b) the entanglement follows a linear scaling after an initial transient spreading from the defect (plotted every 6 Floquet cycles at $T = 4.2$) and (d) the half partition entanglement exhibits a steady linear growth at early times.} 
  \label{fig2_evolution}
\end{figure}


\textit{Heating as quasienergy folding} ---  We first approach the problem by studying the growth of EE in the driven free-fermionic impurity model Eq.~\eqref{eq:floquetdrive_free}. While our results hold for general Floquet drives, for simplicity we restrict our attention to a 2-step drive in both the gapless NN and gapped SSH models. Specifically, this 2-step protocol alternates the impurity parameter $\lambda(t)$ between $1$ (impurity-free chain) and some value $\lambda_0 < 1$, that parameterizes the driving strength. The dynamics are initialized from the half-filling ground state of the corresponding impurity-free models.
For $\lambda_0 = 0.5$, the growth of EE in the gapless NN chain is shown in Figs.~\ref{fig2_evolution}(a,c) for $T<\pi$, and~\ref{fig2_evolution}(b,d) for $T>\pi$. The corresponding plots for the gapped model are provided in the SM \cite{supmat}.
We observe in Fig.~\ref{fig2_evolution}(a) a logarithmic spatial scaling of EE even at late times, accompanied by temporal oscillations clearly seen in Fig.~\ref{fig2_evolution}(c). 
This picture is consistent with local quench dynamics, where quasiparticles are emitted only from the defect, and the growth of EE is subextensive and exhibits revivals for finite systems. After the heating transition (which corresponds to the single particle bandwidth, after which quasienergy folding occurs), EE grows linearly, as generically expected for a driven system. 

A standard diagnostic for heating transitions in Floquet driven systems lies in the spectrum of the single-particle Floquet unitary $U_\text{F}=\mathcal{T}\exp\left(-i\int_0^T \text{d}t H(t)\right)$. 
In the non-heating phase, the eigenvalues of $U_\text{F}$, $u_n = \exp(i\epsilon_{F, n} T)$, are non-degenerate and lie on a unit circle. In this regime, the driven impurity acts as an effectively localized, weak perturbation that fails to induce resonant transitions between distinct bulk eigenmodes.
The transition occurs at $T_\ast = \pi$, when the driving frequency $\omega=2\pi/T$ matches the single-particle bandwidth, causing a $\pi$-gap closure in the spectrum of $U_\text{F}$. 
Beyond this point, previously non-resonant modes become degenerate in quasienergy, satisfying the resonance condition $\epsilon_f - \epsilon_i = n\omega\ (n \in \mathbb{Z})$. 
This enables the drive to hybridize spatially extended modes, producing the volume-law entanglement growth.

For the gapped SSH chain, the same quasienergy folding mechanism applies, and the universal transition at $T_\ast = \pi$ persists (see SM \cite{supmat}). However, the 2-band structure of the SSH model introduces qualitatively new features: the non-heating regime now maintains an area-law entangled structure before transitioning to a volume-law phase within the heating regime, and the interplay between the upper and lower bands generates multiple quasienergy folding points, at both 0- and $\pi$-gaps, leading to an alternating sequence of heating and non-heating phases as $T$ increases (see Fig.~\ref{fig1_impurity}(c)). We present the detailed SSH dynamics, including its dependence on the dimerization $\delta$ and topology, in the End Matter and SM~\cite{supmat}.

The universality of this phenomenon is further strengthened by the fact that the gapless NN chain with a local harmonic drive exhibits the same critical period $T_\ast = \pi$. Notably, this case is analytically tractable: the Floquet Hamiltonian can be obtained in closed form~\cite{PhysRevB.103.L041405} (see SM~\cite{supmat}), with the gap closure at $T_\ast$ verified exactly. We thus conclude that the entanglement transition induced by a local impurity is governed by the onset of quasienergy folding, at which a local periodic drive activates non-local single-particle resonances that hybridize distant modes and drive global heating.


\textit{Energy localization transition} --- While the spectrum of $U_\text{F}$ provides a transparent single-particle picture of the transition, it fails to capture the dynamics at the level of many-body states. Given that the dynamics generated by the Floquet drive remains local below $T_*$, the many-body initial state is expected to mostly overlap with a specific Floquet eigenstate. 
We show that this desired eigenstate can be identified by adopting a recent geometric formulation of Floquet theory~\cite{schindler2025geometricfloquettheory}. This framework resolves the quasienergy gauge ambiguities of the standard Floquet Hamiltonian $H_\text{F}$ by constructing a gauge-invariant ``average energy" operator that possesses a unique ground state, which is precisely the targeted physically relevant state.

The average energy operator is defined via the Floquet eigenstates $|\psi_n[t]\rangle$ of the Floquet Hamiltonian in a given Floquet gauge $t$ (i.e., a choice of time origin, with $H_{\text{F}}[t]$ defined via $U(t+T,t) = e^{-iH_{\text{F}}[t]T}$):
\begin{equation}
\label{eq:averageenergyoperator}
\Theta(T)=\sum_n \theta_n(T)|\psi_n[0]\rangle\langle\psi_n[0]|,
\end{equation}
where the eigenvalues $\theta_n(T)$ are explicitly computed as $\theta_n(T)=\frac{1}{T}\int_0^T \text{d}t \langle \psi_n[t]|H(t)|\psi_n[t]\rangle$, enabling a definitive sorting of Floquet eigenstates free from any gauge ambiguity. 

\begin{figure*}[!htbp]
    \centering
    \includegraphics[width=\linewidth]{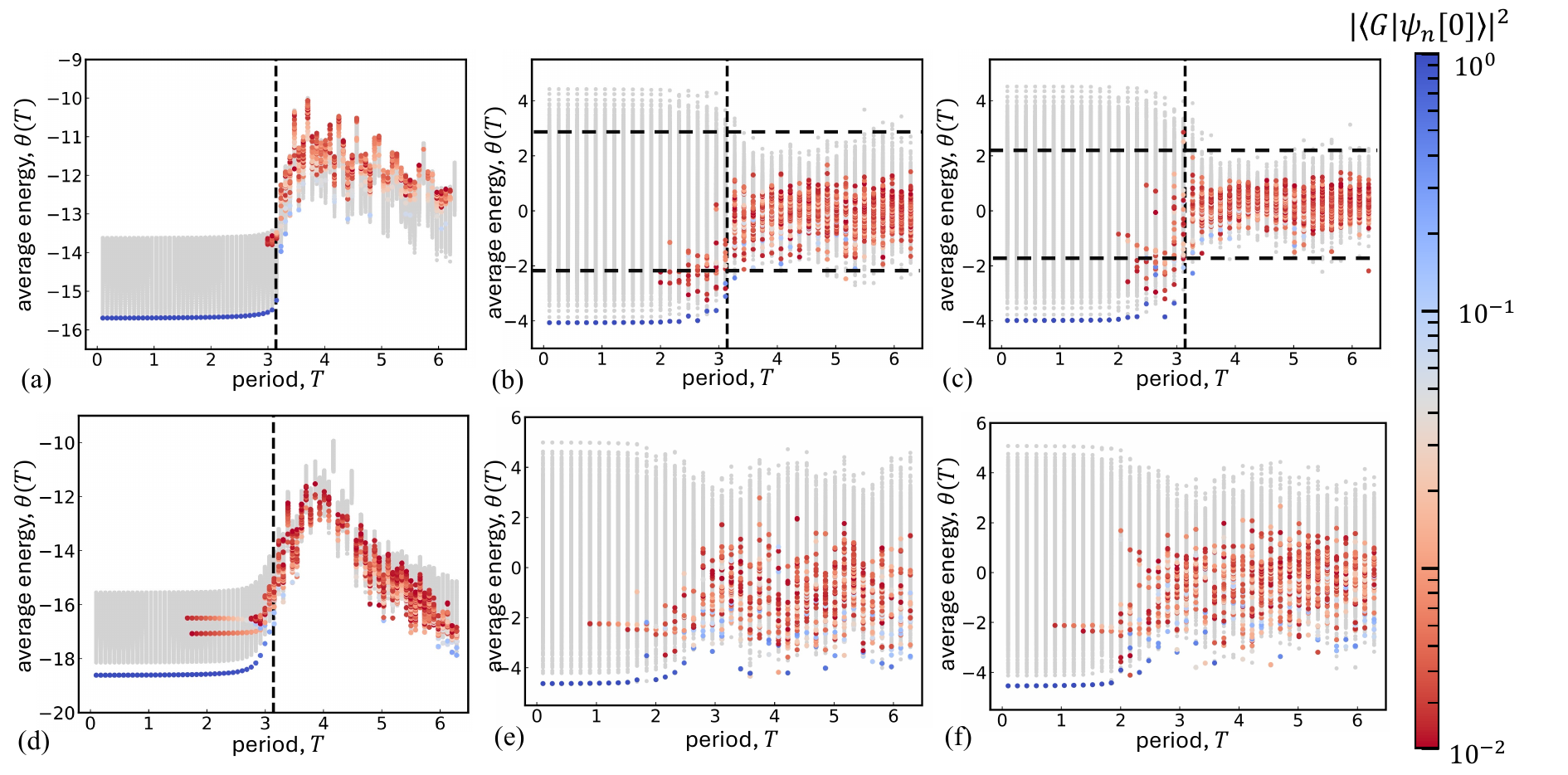}
    \caption{Average Energy spectrum $\theta_n(T)$ as a function of driving period $T$ for a 2-step drive with $\lambda_0=0.5$. Noninteracting limits are shown for (a) the gapless NN model and (d) the gapped SSH model with $\delta=-0.5$, displaying the lowest $10^5$ states with system size $2L=50$. Corresponding spectra in the presence of NN interactions $\Delta=0.1,0.15$ are presented for (b-c) the gapless NN model and (e-f) the gapped SSH model, for system size $2L=14$. The dashed vertical line in (a-d) marks the noninteracting transition point $T_\ast=\pi$, and the dashed horizontal lines in (b-c) mark the bandwidth for $T>\pi$. The color indicates the overlap between the ground state of the infinite frequency effective Hamiltonian $\bar{H} = (1/T)\int_0^T H(t) \mathrm{d}t$ with the ground state of $\Theta(T)$. Grey dots indicate that the overlap is smaller than $0.002$.}
    \label{fig:average_energy}
\end{figure*}

We now illustrate that the entanglement transition across $T_\ast = \pi$ in the driven impurity model corresponds to a breakdown of ``adiabatic continuity'' for $\Theta(T)$. To establish this, we compute the spectrum of Eq.~\eqref{eq:averageenergyoperator} using exact diagonalization (ED) for both gapless and gapped models across $T_\ast$. In both cases, we find that the average energy many-body spectra undergo a localization transition: for any finite periodicity $T<\pi$, the ground state of average energy operator stays close to the ground state at infinite frequency $T\rightarrow0$. 
Therefore, the finite frequency spectrum is adiabatically connected to the infinite frequency spectrum, and the dynamics is effectively captured by the infinite frequency limit, which consists of a single local quantum quench. 
On the other hand, above the critical periodicity, the infinite frequency ground state delocalizes in the Floquet basis (Figs.~\ref{fig:average_energy}(a, d)), and the resulting dynamics is no longer expected to be local. 
Notably, the infinite frequency effective Hamiltonian $\bar{H}$ differs from the impurity-free Hamiltonian only by a local static defect, so its ground state approximates our initial state. This confirms that the ground state of $\Theta(T)$ is precisely the physically relevant Floquet eigenstate. 

This many-body diagnosis naturally extends to the interacting cases, obtained by adding NN interactions $\Delta\sum_j n_jn_{j+1}$ to Eq.~\eqref{eq:floquetdrive_free}. 
For the gapless NN model (Figs.~\ref{fig:average_energy}(b-c)), adding small interactions does not significantly alter the above picture: the ``adiabatic continuity'' to the infinite frequency persists over a wide frequency range before eventually delocalizing above a critical frequency threshold. This is accompanied by photon resonances which drastically reduce the bandwidth upon increasing interaction strength. 
Conversely, the interactions have a more dramatic effect on the gapped model (Figs. \ref{fig:average_energy}(e-f)). While the Floquet ground state still dominates the dynamics within a finite frequency window, the delocalization threshold drifts away from the noninteracting value. Due to strong finite-size effects, a definitive extrapolation of the dynamical phase diagram in the interacting SSH model remains computationally challenging.

Furthermore, we also investigate the spatial structure of $\Theta(T)$ in both models in the End Matter. We find that in the non-heating phase, the operator $\Theta(T)$ maintains a local structure, while it develops emergent long-range matrix elements in the heating phase, consistent with the observed entanglement dynamics.


\textit{MPS simulations} --- The average spectra of the interacting gapless model (Figs.~\ref{fig:average_energy}(b-c)) suggests that the non-heating window remains robust in the presence of weak interactions. 
We now employ MPS calculations in order to better understand the entanglement dynamics of this driven \textit{interacting} model ($\Delta \ne 0$).
This model maps to a locally driven XXZ chain via Jordan Wigner transformation, where the NN interaction strength $\Delta$ is exactly the anisotropy in spin language.
As in the noninteracting setup, we initialize the dynamics in the ground state of the impurity-free interacting gapless chain, then apply the 2-step drive with $\lambda_0 =0.5$.

\begin{figure}[!t]
    \centering
    \includegraphics[width=\linewidth]{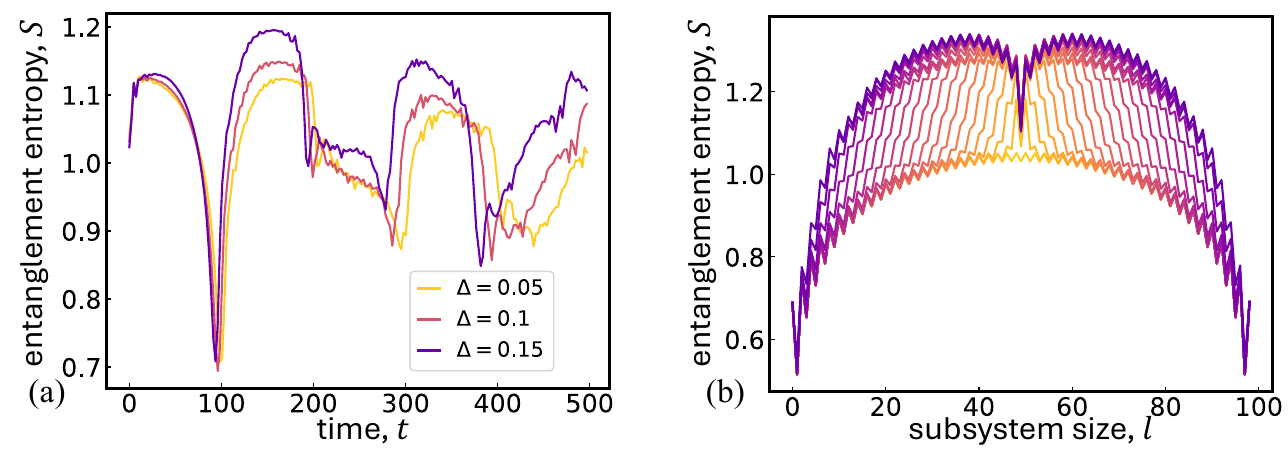}
    \caption{Entanglement evolution of an interacting gapless chain of system size $2L=100$ subject to 2-step drive with $\lambda_0=0.5$ within the non-heating regime. (a) In the non-heating regime ($T=2.4$), the half-partition entanglement remains bounded after a few hundred Floquet cycles across different $\Delta$ values, akin to the noninteracting case. Notably, the revival period now develops a dependence on the interaction strength $\Delta$. 
    (b) For $\Delta=0.1$, the EE scales logarithmically over space for intermediate times (plotted every cycle, from yellow to purple). The MPS parameters used are: an initial DMRG bond dimension of $\chi_D = 100$, which determines the ground state and a TEBD bond dimension of $\chi_T=400$ for $\Delta=0.05, 0.1$ and $\chi_T=500$ for $\Delta = 0.15$.} 
    \label{fig:MPS_evo}
\end{figure}

Our MPS simulations reveal that the non-heating behavior observed in the noninteracting model persists in the presence of interactions up to numerically accessible timescales, i.e., several hundreds of Floquet cycles. 
As can be seen in Fig.~\ref{fig:MPS_evo}(a), EE does not exhibit significant heating at least for a timescale of $\mathcal{O}(L)$, at a finite driving frequency which lies within the non-heating phase of the noninteracting setup. 
Our simulations clearly indicate the existence of a finite non-heating regime that differs from typical heating behavior in driven systems. In particular, EE still features partial revivals, whose periodicity explicitly depends on $\Delta$. This phenomenon can be explained by an effective local quench picture where quasiparticles are emitted from the impurity, as detailed in the SM~\cite{supmat}. 
Furthermore, within this stable regime, the EE retains a logarithmic scaling after a transient spreading from the impurity (see Fig.~\ref{fig:MPS_evo}(b)), as opposed to the expected volume law scaling in driven interacting nonintegrable models. 

While one may naturally expect the non-heating dynamics observed here to be limited to a finite timescale, and for the system to eventually thermalize, we stress that the mechanism at work here is distinct from usual prethermalization~\cite{yu2019discrete,khemani2019review,bukov2021pre,romero2022fractional,mukherjee2024emergent,yan2024pre}. Prethermalization, in the absence of disorder and away from high-frequency expansions, typically relies on emergent quasi-conserved quantities enabled by the global drive.
On the other hand, the locally driven interacting gapless model appears to be stable for a continuous range of frequencies far away from the high frequency expansion, and does not rely on any particular symmetry due to the impurity. Note that similar conclusions have recently been reported for locally driven interacting quantum point contacts~\cite{dudinets2024fermionictransportdrivenquantum}, where the system remains insulating for infinitely large times (but for small system sizes reachable by ED) for a continuous range of frequencies above the critical frequency. 


\textit{Outlook} --- 
We have uncovered an entanglement transition between local and global EE growth, tuned by the period of a Floquet-driven impurity. In the noninteracting limit, this coincides with the onset of quasienergy folding and thus a breakdown of high-frequency adiabatic continuity. The gapless NN chain shows a single transition; the gapped SSH chain is richer, folding at both $0$ and $\pi$ gaps to give an alternating sequence of heating and non-heating phases. We expect that this mechanism is generic to locally driven free-fermion chains. With interactions, we cannot rigorously establish a transition, but our gapless simulations give strong evidence for a robust non-heating regime with persistent EE oscillations, far from the high frequency regime and without spatial disorder. The interacting SSH chain remains an open problem due to strong finite-size effects.

Our findings are amenable to experimental verification in state-of-the-art cold-atom setups. Single-site, time-dependent modulations have been demonstrated~\cite{Krinner2017,Trisnadi_2022,Young_2022} and techniques using acousto-optical deflectors allow for precise temporal control of local on-site energies in optical lattices~\cite{Nixon_2024}. Crucially, the essential physics in our setup does not depend on the specific form of the local modulation, making it robust to experimental imperfections like leakage to neighboring sites. While direct measurement of EE is challenging, the transition could be witnessed through correlators that probe transport or energy growth across the system. Furthermore, our proposed mechanism, which allows for tuning between regimes in a single experiment by locally driving the system, offers a versatile platform to study impurity dynamics in cold atom and circuit QED experiments.

Beyond the unitary dynamics explored in the main text, we also show in the SM~\cite{supmat} that the transition between logarithmic and volume-law entanglement scaling persists even when the impurity is non-Hermitian. This setup can be interpreted as a periodically monitored system at a single site in the no-click limit~\cite{Stefanini_2024}. We anticipate that a rich criticality underlies this non-unitary entanglement transition, given that non-Hermitian impurity models have recently been connected to non-unitary conformal field theories~\cite{li2025impurityinducednonunitarycriticality}. Broadly speaking, our work opens a new avenue for exploring novel critical phenomena induced by local Floquet driving.


\begin{acknowledgments}\textit{Acknowledgments} --- We thank Marin Bukov, Pieter Claeys, Paul Schindler, Konrad Viebahn, and Hongzheng Zhao for fruitful discussions. We also thank the anonymous referees for their constructive comments.
We acknowledge the use of the software packages \texttt{QuSpin}~\cite{10.21468/SciPostPhys.2.1.003} for ED calculations, and \texttt{TenPy}~\cite{10.21468/SciPostPhysLectNotes.5} for DMRG and TEBD numerical simulations. 
B.L. acknowledges financial support from the Swiss National Science Foundation (Postdoc.Mobility Grant No. 214461). 
S.R. is supported by the National Science Foundation under Award No.\ DMR-2409412.
This work was supported in part by the Sivian Fund and Paul Dirac Fund at the Institute for Advanced Study, the U.S. National Science Foundation under Grant No. PHY-2309135 to the Kavli Institute for Theoretical Physics (KITP), and the U.S. Department of Energy, Office of Science, Office of High Energy Physics under Grant No. DE-SC0009988. This material is based upon work supported by the U.S. Department of Energy, Office of Science, Office of Advanced Scientific Computing Research via the Exploratory Research for Extreme Scale Science (EXPRESS) program under Award Number DE-SC0026216. A.P. thanks KITP for its hospitality during the “Noise-robust Phases of Quantum Matter” program, during which part of this work was completed. 
Z.-X.L. thanks the Max Planck Institute for the Physics of Complex Systems for its support and hospitality, during which significant progress was made on this work. 
S.R. would like to thank the Isaac Newton Institute for Mathematical Sciences, Cambridge, for their support and hospitality during the program ''Quantum field theory with boundaries, impurities, and defects," where work on this paper was undertaken. 
\end{acknowledgments}
\textit{Disclaimer ---} This report was prepared as an account of work sponsored by an agency of the United States Government. Neither the United States Government nor any agency thereof, nor any of their employees, makes any warranty, express or implied, or assumes any legal liability or 	responsibility for the accuracy, completeness, or usefulness of any information, apparatus, product, or process disclosed, or represents that its use would not infringe privately owned rights.  Reference herein to any specific commercial product, process, or service by trade name, trademark, manufacturer, or otherwise does not necessarily constitute or imply its endorsement, recommendation, or favoring by the United States Government or any agency thereof. The views and opinions of authors expressed herein do not necessarily state or reflect those of the United States Government or any agency thereof.



%

\onecolumngrid 
\section*{End Matter}

\twocolumngrid
\textit{Local driven impurity in SSH model} --- We now demonstrate that the heating transition at $T_\ast = \pi$ persists in gapped model, where the phase diagram is significantly enriched. We consider the SSH model, defined by taking $t_i = 1 - (-1)^i \delta$ in \eqref{eq:floquetdrive_free}, subject to a 2-step driven impurity. Here $\delta$ is the dimerization parameter ($|\delta| < 1$), where $\delta > 0$ and $\delta<0$ characterize the non-topological and topological phases, respectively. 

In \cite{supmat}, we simulate the EE evolution initialized from the half-filling ground state of the homogeneous model for both the topological and non-topological phases, confirming that the heating transition at $T_\ast = \pi$ remains robust in the presence of a bulk gap. As established in the main text, this universal transition occurs when the driving frequency matches the total bandwidth. 
Additionally, by checking the evolution of the EE scaling at stroboscopic times, we discover that the gapped structure of the SSH chain critically reshapes the nature of the entanglement transition: while the locally driven gapless model exhibits a transition from log to volume-law scaling, the SSH model manifests a transition from area to volume law entanglement.

\begin{figure*}[!htbp]
    \centering
    \includegraphics[width=\linewidth]{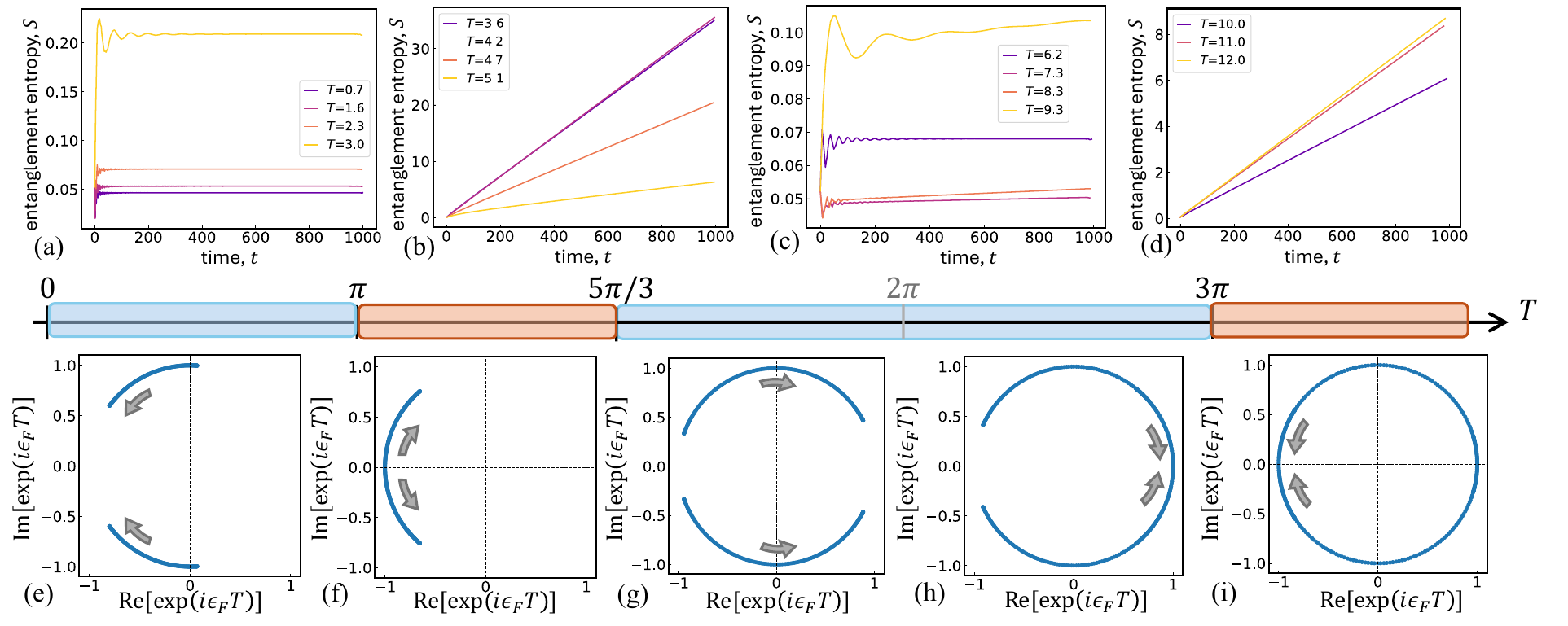}
    \caption{The phase diagram of the non-topological SSH model with $\delta = 0.6$ and system size $2L=400$, under a 2-step drive with $\lambda_0 = 0.5$. The system exhibits alternating transitions between non-heating and heating phases at $T_\ast \in \{\pi, 5\pi/3, 3\pi\}$. (a)-(d) present the half-partition EE evolution for each phase. (e)-(i) demonstrate the corresponding spectrum of the Floquet unitary in each regime, with $T = 2.5, 4.0, 5.8, 9.0$ and $10.0$. Notably, while a 0-gap closure occurs at $T = 2\pi$, heating remains suppressed in the system. This ``dark" resonance is attributed to the vanishing of the relevant Fourier components in the 2-step drive, which prevents the coupling required to trigger many-body heating despite the quasienergy degeneracy.}
    \label{fig:SSH_phase_nontopo}
\end{figure*}

The 2-band structure of the SSH model further enriches the dynamical phase diagram. When $|\delta|$ is large enough, we observe multiple transition points with the system alternating between non-heating and heating phases as the driving period $T$ increases (see Fig.~\ref{fig:SSH_phase_nontopo}). 
These transitions are physically driven by resonances between the lower and upper bands, occurring when the drive induces a closure of either the 0- or $\pi$-quasienergy gap.
Specifically, at $\delta = 0.6$ (non-topological phase) with a two-step driven impurity, we identify a sequence of transition points at $T_\ast \in \{\pi, 5\pi/3, 3\pi\}$. 

Furthermore, we observe distinct phase boundaries between the topological and non-topological regimes \cite{supmat}. This discrepancy is attributed to the presence of topological edge modes localized at the impurity site, which mediate additional resonance channels with the bulk states that are absent in the non-topological case. 
The structure of the phase diagram also depends critically on $|\delta|$, with the number of transition points increasing monotonically with $|\delta|$. A comprehensive study of the full phase diagram and the underlying Floquet resonance physics are left for future works. 

\textit{Average energy operator} --- As illustrated above, the spectrum of $U_\text{F}$ provides an intuitive picture of the heating transitions in the driven impurity model. Nevertheless, it's desirable to establish a many-body perspective to characterize the transition. Given Floquet eigenstates are transitionless under stroboscopic evolution, the presence of non-heating phases implies that our initial state is ``close'' to a specific many-body Floquet eigenstate. However, the standard Floquet Hamiltonian fails to identify which eigenstate is physically relevant. This motivates the application of the Average Energy operator, recently proposed in~\cite{schindler2025geometricfloquettheory}. 

\begin{figure*}[!htbp]
    \centering
    \includegraphics[width=0.8\linewidth]{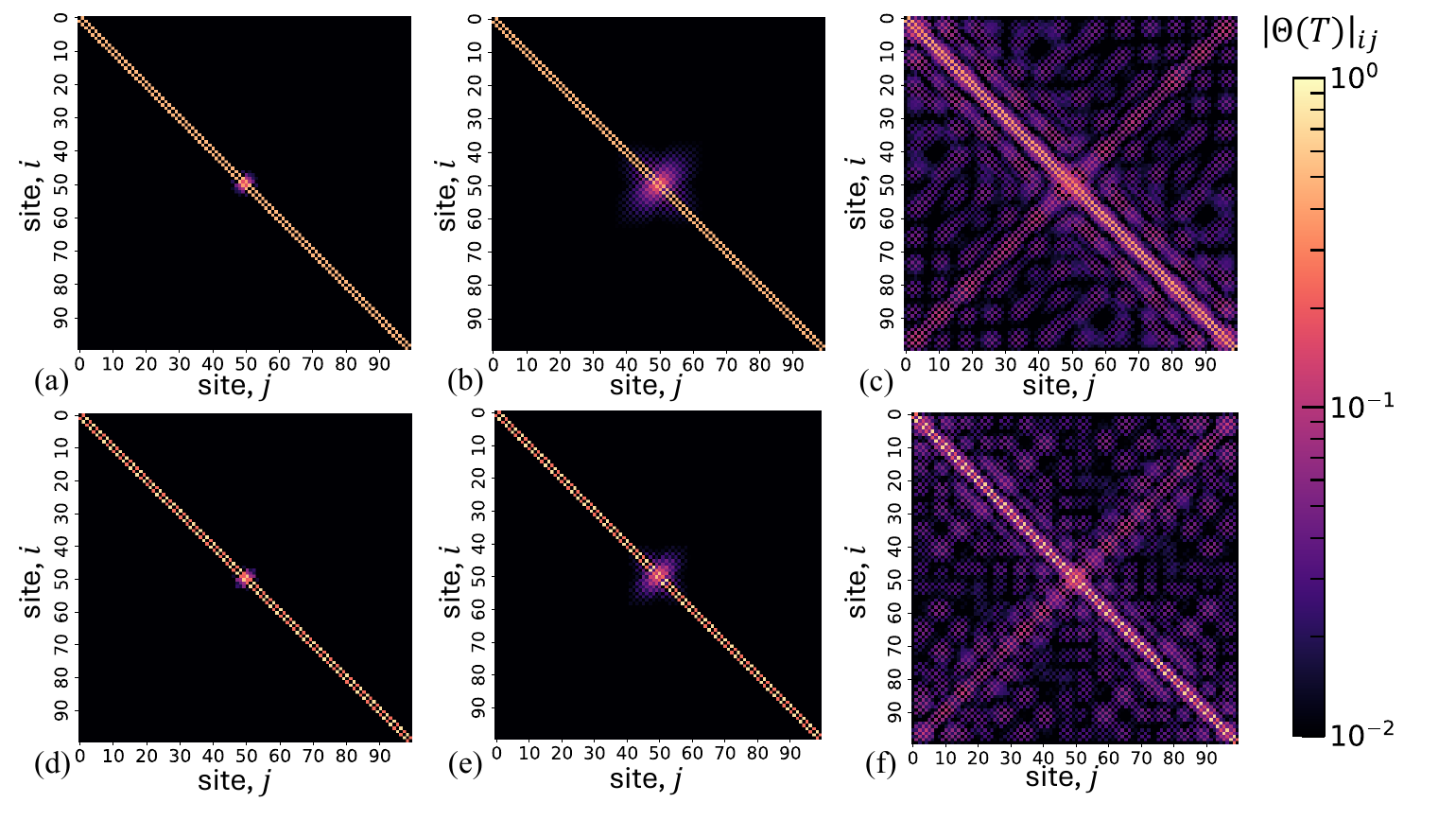}
    \caption{Spatial structure of the single-particle average energy operator $\Theta(T)$ in (a)-(c) the gapless NN chain and (d)-(f) the gapped SSH chain ($\delta = -0.5$)  with system size $2L=100$ under a 2-step driven impurity with $\lambda_0=0.5$. The color indicates the norm of the matrix element $|\Theta(T)|_{ij}$. The operator exhibits a universal behavior across both gapless and gapped models: (a) (d) in the non-heating regime ($T=2.8$), $\Theta(t)$ remains spatially local; (b) (e) at the critical point $(T = \pi)$, off-diagonal structures emerge from the impurity site; and (c) (f) in the heating regime ($T = 3.3$), it develops a significantly long-range coupling pattern.}
    \label{fig:AE_heatmap}
\end{figure*}

To start with, we recap the standard Floquet formalism and explain its limitation. 
For any periodically driven system $H(t) = H(t+T)$, the time evolution unitary is factorized into the stroboscopic and micromotion components~\cite{viebahn2020introduction}:
\begin{equation}
   U(t) = \mathcal{T}\exp\left(-i\int_0^t H(\tau)\mathrm{d}\tau\right)= P(t) \exp(-it H_\text{F}[0]) ,\quad  
   \label{eq:Floquet_def}
\end{equation}
where $H_\text{F}[t]$ is the Floquet Hamiltonian at Floquet gauge $t$, and $P(t)$ is the time-periodic micromotion operator. They are related to $H(t)$ through $H(t) = H_\text{F}[t] + i\partial_t P(t)  P^\dagger(t)$, while the choice of $H_\text{F}$ and $P(t)$ is in principle non-unique. This stems from the ambiguities of quasienergies $\epsilon_n \to \epsilon_n + 2\pi m /T, m \in \mathbb{Z}$, which prevents a definitive sorting of states and hinders the physical interpretation of the Floquet Hamiltonian. The lack of a unique state ordering makes it difficult to map an initial state to its corresponding Floquet eigenstate. 

To address these challenges, we adopt an alternative formulation of Floquet theory based on a geometric aspect~\cite{schindler2025geometricfloquettheory}. This approach interprets the Floquet problem as a counterdiabatic driving, generating transitionless evolution of Floquet eigenstates $\{|\psi_n[t]\rangle\}$. In this way, the time evolution unitary is uniquely factorized as,
\begin{equation}
    U(t, 0) = \mathcal{W}(t) \exp(-it \Theta(t)) ,
\end{equation}
where $\mathcal{W}(t)$ is the Wilson line operator describing the geometric part of the evolution (see SM~\cite{supmat}),
and $\Theta(t)$ is the average energy operator, representing the dynamical contribution, 
\begin{align}
    &\Theta(t) =\sum_n \theta_n(t)| \psi_n [0]\rangle \langle  \psi_n[0]|, \nonumber \\
    &\theta_n(t) = \frac{1}{t} \int_0^t \langle \psi_n[\tau] |H(\tau) | \psi_n[\tau]\rangle \mathrm{d}\tau 
\end{align}
Crucially, at stroboscopic times, the average energy operator $\Theta(T)$ is a gauge-invariant of motion. In contrast to the Floquet Hamiltonian, this unique decomposition enables an unambiguous sorting of the Floquet eigenstates $\{|\psi_n[t]\rangle\}$ according to their average energies $\theta_n(T)$. 
This, in turn, enables the unambiguous identification of a Floquet ground state at any driving period $T$. 
As demonstrated in~\cite{schindler2025geometricfloquettheory} and supported by our findings, the average energy spectrum $\theta_n(T)$ serves as a sensitive probe for dynamical transitions, with the Floquet ground state corresponding precisely to the physically relevant state.

We now apply this construction to the 2-step driven impurity in both the gapless XXZ chain and the gapped (interacting) SSH model to identify spectral signatures of the heating transition near $T_\ast = \pi$.
We start by considering the Floquet unitary operator, $U_\text{F}[0]=e^{-i H_0 T/2}e^{-i H_1 T/2}$ with driving period $T$, where $H_0$ and $H_1$ denote the Hamiltonians in Eq.~\eqref{eq:floquetdrive_free} with $\lambda(t)=1$ and $\lambda(t)=\lambda_0$, respectively. By evaluating the average energy operator at stroboscopic times, we obtain (see \cite{supmat}), 
\begin{equation}
\Theta(T) =\frac{1}{2}\sum_n  \langle\psi_n[0]|H_0 +  H_1 |\psi_n[0] \rangle | \psi_n[0] \rangle  \langle\psi_n[0]|.
\end{equation}
In the noninteracting limit, $\Theta(T)$ takes a quadratic form, which can be fully represented by a $2L\times 2L$ single particle matrix. We analyze the spatial distribution of $\Theta(T)$ in Fig.~\ref{fig:AE_heatmap}. In the non-heating phase $(T < \pi)$, $\Theta(T)$ is dominated by nearest-neighboring couplings, closely resembling the static Hamiltonian. Critically, at the transition point $T=\pi$, we observe the emergence of long-range couplings starting from the impurity site. As the system enters the heating phase, these long-range couplings spread across the system, most prominently along the anti-diagonal. 

This formalism extends naturally to the interacting regime. We compute the average energy $\theta_n(T)$ for both gapless and gapped models around the universal transition point $T_\ast = \pi$, with which the many-body Floquet eigenstates are sorted. 
Our numerical results demonstrate that, as long as $T<\pi$, the Floquet ground state, defined by the minimum average energy $\theta_0(T)$, maintains a near-unity overlap with the ground state of the time-averaged Hamiltonian $(H_0+H_1)/2$, the infinite frequency ground state $|G\rangle$. However, this overlap drops sharply to zero and $|G\rangle$ delocalizes across the Floquet eigenbasis once the system enters the heating phase $T>\pi$ (see Fig.~\ref{fig:average_energy}). This delocalization behavior near $T_\ast = \pi$ indicates a breakdown of the adiabatic continuity at finite frequency, consistent with the onset of the heating dynamics revealed in our findings, thereby provides a robust many-body diagnostic for the entanglement dynamics in our model.


\onecolumngrid 
\clearpage
\makeatletter 

\begin{center}   
	\textbf{\large Supplementary Material for ``Local-to-Global Entanglement Dynamics by Periodically Driving Impurities"}\\
	[1em]
	Zhi-Xing Lin$^1$, Abhinav Prem$^{2,3}$, Shinsei Ryu$^{1}$, and Bastien Lapierre$^{1,4}$ \\[.1cm]
	{\itshape \small ${}^1$Department of Physics, Princeton University, Princeton, New Jersey, 08544, USA \\ 
	${}^2$Physics Program, Bard College, 30 Campus Road, Annandale-on-Hudson, New York 12504, USA\\
    ${}^3$School of Natural Sciences, Institute for Advanced Study, Princeton, New Jersey 08540, USA\\
    ${}^4$Philippe Meyer Institute, Physics Department, École Normale Supérieure (ENS), Université PSL, 24 rue Lhomond, F-75231 Paris, France}\\
	(Dated: June 19, 2026)\\[1cm]
\thispagestyle{titlepage} 
\end{center} 	

\renewcommand{\thefigure}{S\arabic{figure}}
\setcounter{figure}{0}

\appendix
\setcounter{secnumdepth}{2} 

\addcontentsline{toc}{section}{Supplementary Material}
This Supplemental Material contains a number of Appendices with technical details supporting the results presented in the main text: 
\begin{enumerate}
    \item Appendix \ref{APP:driven_SSH} delineates the entanglement dynamics of the SSH model subject to a locally driven impurity, revealing a richer dynamical phase diagram structure compared to the gapless case.
    \item Appendix~\ref{App:geometricfloquettheory} establishes the formal framework of the Kato Hamiltonian and average energy operator and presents the application of these concepts to our models.
    \item In Appendix~\ref{App:mpscalculations}, we reveal that the non-heating phase is characteristic of an effective local quench, exhibiting periodic partial revivals and logarithmic temporal growth.
    \item Appendix~\ref{App:nonhermitianfloquet} demonstrates that the local-to-global entanglement dynamics persists even for a non-Hermitian driven impurity, which has an enriched phase diagram with an additional spontaneous parity-time symmetry breaking transition. 
    \item Appendix~\ref{App:universality} demonstrates the universality of the entanglement transition at $T_\ast = \pi$ by showing that it persists across qualitatively different impurity configurations, including hopping, double, and single impurities.
    \item Appendix \ref{APP:MPS_backup} details the numerical protocols of our MPS simulation, justifying our parameter choices and establishing the numerical stability of our results. 
    \item  In Appendix~\ref{App:floquetham}, we include the analytical derivation of the Floquet Hamiltonian and its spectrum for a harmonically driven impurity. 
    \item In Appendix~\ref{App:schriefferwolff}, we use the Floquet Hamiltonian derived in Appendix~\ref{App:floquetham} to demonstrate that the effective low-energy dynamics in the high-frequency limit is governed by local perturbations around the impurity.
\end{enumerate}


\section{Entanglement Dynamics of SSH Model with Local Drives}
\label{APP:driven_SSH}

This Appendix details the entanglement dynamics of a SSH chain under a local driven impurity. The Hamiltonian writes,
\begin{equation}
    H(t) = -\frac{1+\delta}{2} \sum_{j=1}^{L}\left(c^\dagger_{2j} c_{2j-1}   +  h.c.\right) -\frac{1-\delta}{2}\left(\sum_{j=1}^{L/2-1}+\sum_{j=L/2+1}^{L-1}\right) \left(c^\dagger_{2j+1} c_{2j}  +  h.c.\right) + V(t)
    \label{floquetdrive_SSH}
\end{equation}
where
\begin{equation}
    V(t) = \frac{1-\delta}{2} \left( c_{L}^\dagger \ c_{L+1}^\dagger \right)
    \left(
    \begin{array}{cc}
       \sqrt{1-\lambda^2(t)}  &  -\lambda(t) \\
       -\lambda(t)  &  -\sqrt{1-\lambda^2(t)}
    \end{array}
    \right) \left(\begin{array}{c}
         c_{L}  \\
         c_{L+1} 
    \end{array}
    \right),
\end{equation}
The setup is equivalent to taking $t_i = 1-(-1)^i\delta$ in \eqref{eq:floquetdrive_free}. 
Our motivation to investigate such a system comes from several perspectives. 
First, the gapped band structure offers a richer landscape for many-body resonance channels, which can be selectively closed or opened by tuning the driving periods $T$. 
Second, the SSH model serves as a natural platform to probe the interplay between topological phases and dynamical heating, specifically, whether the topological and non-topological regimes have distinct signatures on the entanglement dynamics.

\begin{figure}[!htbp]
    \centering
    \includegraphics[width=\linewidth]{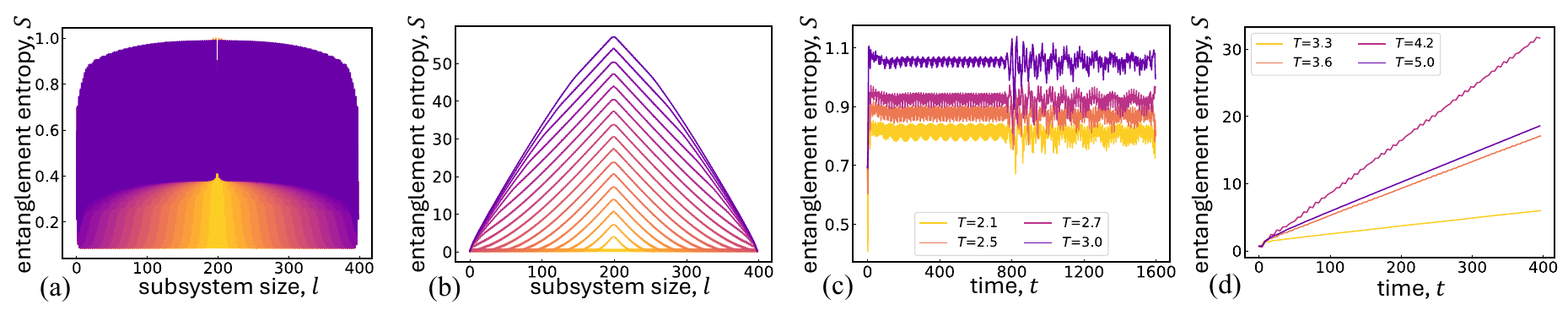}
    \caption{The entanglement dynamics of a topological SSH chain with $\delta=-0.5$ and system size $2L=400$, under a 2-step driven impurity with $\lambda_0=0.5$. The universal dynamical transition around $T_\ast = \pi$ remains robust with the presence of the two-band structure. (a)(b)The entanglement scaling (plotted every 10 cycles, from orange to purple), experiencing a transition from (a) an area law in the non-heating phase at $T=2.5$, to (b) a volume law in the heating phase $T=4.2$. (c)(d) The half-partition entanglement evolution across different driving periods $T$ in both regimes.}
    \label{fig:SSH_evo}
\end{figure}

To begin with, we establish that the entanglement dynamics around $T_\ast = \pi$ remains robust across both gapless and gapped models (see Fig.~\ref{fig:SSH_evo}). 
The transition is different in one aspect: in gapped models like SSH chains, initializing the dynamics from the half-filling ground state of a homogeneous SSH chain provides an area-law entangled state, differing fundamentally from the log-law entangled ground state in the gapless system. Therefore, the entanglement features a transition from an area law to a volume law. 

However, we further observe that the phase diagram in gapped systems has a much richer structure away from $T_\ast = \pi$. 
As established in the main text, the onset of heating is attributed to the folding of quasienergy spectrum, where the resonance condition $E_f - E_i = n\omega$, ($n\in \mathbb{Z}$ with $\omega$ being the driving frequency) allows the external driving to trigger resonant hopping between distant energy levels. This picture clearly highlights how the presence or absence of a band gap plays a critical role in the dynamical transitions: in a gapless system, once the driving frequency $\omega$ falls below the band width $W$, quasienergy folding persists for all $\omega < W$, driving the system into a permanent heating regime. 
Nevertheless, this story becomes significantly more complex with the presence of a band gap. While the initial transition to heating occurs when $\omega$ matches $W$, a further decrease of $\omega$ can cause it to fall below the band gap. At this point, the many-body resonance channels are ``frozen" by the band gap, inducing the reentrance into the non-heating phase. This interplay between the drive frequency and the spectral gaps leads to a sequence of reentrant transitions as the driving period $T$ varies, a cycle that persists until the frequency becomes small enough to sustain permanent many-body heating. 
\begin{figure}[!htbp]
    \centering
    \includegraphics[width=\linewidth]{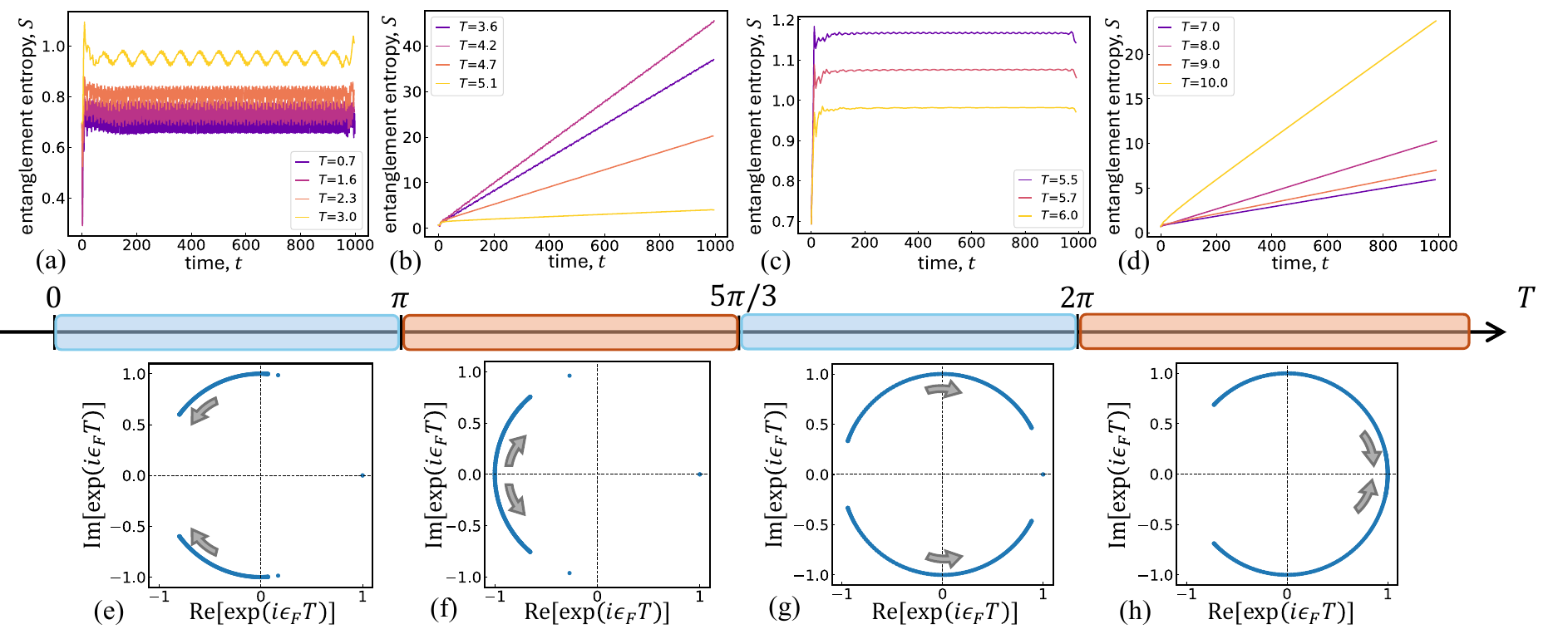}
    \caption{The phase diagram of the topological SSH model with $\delta=-0.6$ and system size $2L=400$, under a 2-step drive with $\lambda_0=0.5$. Similar to the non-topological case, the system exhibits an alternating transition pattern, but at different $T_\ast \in \{\pi, 5\pi/3, 2\pi\}$. (a)-(d) present the half-partition EE evolution for each phase. (e)-(h) demonstrate the corresponding Floquet unitary spectrum in each regime, with $T=2.5, 4.0, 5.8, 6.5$. Each transition is mediated by either a 0- or $\pi$- gap closure, implying the onset of many-body resonance channels.}
    \label{fig:SSH_phase_topo}
\end{figure}

Numerical simulation of the entanglement dynamics (see Fig.~\ref{fig:SSH_phase_nontopo} and Fig.~\ref{fig:SSH_phase_topo}) confirm that this reentrant transition pattern emerges with sufficiently large $|\delta|$. Crucially, we observe a distinction in phase diagrams between the topological and non-topological regimes. For a specific value $|\delta|=0.6$, both systems share the first two transition points at $T_\ast = \pi$ and $T_\ast = \pi/|\delta|=5\pi/3$. Following the previous arguments, the first transition occurs when the driving frequency $\omega$ matches the total bandwidth $W$. The second one takes place as $\omega$ drops below the band gap, while its higher order harmonics exceed $W$. 
Their divergence manifests at the last transition point: mechanically, a transition at $T_\ast = 2\pi$ requires the second harmonic $2\omega$ to resonate with the total band width. 
However, because even harmonics are strictly absent from the Fourier spectrum of our 2-step drive protocol, this resonance cannot be activated linearly and requires a higher-order, two-photon process. 
In the non-topological regime (Fig.~\ref{fig:SSH_phase_nontopo}), this two-photon channel is heavily suppressed by the bulk band gap, yielding the $2\pi$ resonance ``dark'' and deferring the next heating phase until the third harmonic satisfies $3\omega = W$ at $T_\ast = 3\pi$. 
In contrast, in the topological regime (Fig.~\ref{fig:SSH_phase_topo}), the in-gap topological states localized around the impurity serve as ideal virtual intermediate states, providing the resonance channel for $2\omega$ and facilitating the heating transition at $T_\ast = 2\pi$. A more qualitative investigation into the microscopic details and selection rules governing these higher-order resonance processes is reserved for future work. 

\section{Kato Hamiltonian and the Average Energy operator}
\label{App:geometricfloquettheory}

In this Appendix, we provide a detailed overview of the Average Energy operator framework proposed in Ref.~\cite{schindler2025geometricfloquettheory}, followed by the techinical details about the calculation of many-body average energy spectrum in Fig.~\ref{fig:average_energy}. 

\subsection{Brief review of geometric Floquet theory}

As discussed in the end matter, the average energy operator formalism interprets the periodic driven problem as a counterdiabatic driving, generating transitionless evolution of the instantaneous Floquet eigenstates. 
In this representation, the time-periodic Hamiltonian $H(t)$ can be uniquely decomposed as a control Hamiltonian $H_K(t)$ that determines the adiabatic trajectory of the Floquet states, and an adiabatic parallel-transport gauge potential $\mathcal{A}_K(t)$:
\begin{equation}
   \quad  H(t) = H_K(t) + \mathcal{A}_K(t)
   \label{eq:Kato_decomp}
\end{equation}
where these two components are explicitly defined as,
\begin{equation}
    H_K(t) = \sum_n \epsilon_{K, n}(t)\mathcal{P}_n(t), \quad \epsilon_{K, n}(t) = \mathrm{Tr}[\mathcal{P}_n(t) H(t)]
    \label{eq:Kato_def}
\end{equation}
and
\begin{equation}
    \mathcal{A}_K(t) = -\frac{1}{2} \sum_n [\mathcal{P}_n(t), i\partial_t \mathcal{P}_n(t)]
\end{equation}
with $\mathcal{P}_n(t) = | \psi_n [t]\rangle \langle  \psi_n[t]|$ being the projector of the Floquet eigenstates $|\psi_n[t] \rangle$. 

Correspondingly, the Floquet evolution operator can be factorized as,
\begin{equation}
    U(t, 0) = \mathcal{W}(t) \exp(-it \Theta(t)) ,
\end{equation}
where $\mathcal{W}(t)$ is the Wilson line operator describing the geometric part of the evolution,
\begin{equation}
    \mathcal{W}(t) = \mathcal{T}\exp(-i \int_0^t \mathcal{A}_K(\tau) \mathrm{d}\tau)
\end{equation}
and $\Theta(t)$ is the average energy operator, contributing to the dynamical part, 
\begin{equation}
    \Theta(t) =\sum_n \theta_n(t)| \psi_n [0]\rangle \langle  \psi_n[0]|, \quad  \theta_n(t) = \frac{1}{t} \int_0^t \epsilon_{K, n}(\tau) \mathrm{d}\tau 
\end{equation}
Importantly, the average energy operator at stroboscopic times, $\Theta(T)$, is an invariant of motion. The uniqueness of Kato decomposition allows for an unambiguous sorting of the Floquet eigenstates $\{|\psi_n[t]\rangle\}$ according to their average energies $\theta_n(T)$. 
This, in turn, enables the unambiguous identification of a ground state at any drive frequency, which is absent in the standard the Floquet Hamiltonian formalism due to the quasienergy ambiguity. 

\subsection{Numerical calculation of Kato spectrum}

In our 2-step driven impurity model, we aim to numerically compute the average spectrum $\theta_n(T)$ using ED, and observe if distinct spectral features emerge across the transition (see Fig.~\ref{fig:average_energy}). In particular, we characterize a breakdown of adiabatic continuity between the infinite frequency ground state $|G\rangle$ and the Floquet ground state at finite frequency near the critical point $T_\ast=\pi$.

To this end, we start from our Floquet unitary operator, $U_\text{F}[0]=e^{-i H_0 T/2}e^{-i H_1 T/2}$ with driving period $T$, where $H_0$ and $H_1$ denote the Hamiltonian in Eq.~\eqref{eq:floquetdrive_free} with $\lambda(t)=1$ and $\lambda(t)=\lambda$, respectively. 
Diagonalization of $U_\text{F}[0]$ yields the Floquet eigenstates $\{|\psi_n[0]\rangle\}$. The Floquet unitary $U_\text{F}[t]$ with gauge $t\in[0,T]$ is related to $U_\text{F}[0]$ through a unitary transformation,
\begin{equation}
U_\text{F}[t]=\begin{cases}
e^{-iH_1 t}U_\text{F}[0]e^{i H_1 t}, & \quad t\in[0,T/2), \\
e^{i H_0 (T-t)} U_\text{F}[0] e^{-i H_0 (T-t)}, &\quad t\in[T/2,T).
\end{cases}
\end{equation}
This relation enables us to construct the Floquet eigenstates at any $t$ by rotating those at $t=0$, and the instantaneous Kato spectrum is then given by,
\begin{equation}
\epsilon_{K, n}(t) = \langle\psi_n[t]| H(t) | \psi_n[t] \rangle =\begin{cases}
\langle\psi_n[0]| H_1 |\psi_n[0] \rangle, &\quad t\in[0,T/2), \\
\langle\psi_n[0]| H_0 |\psi_n[0] \rangle,  &
\quad t\in[T/2,T),
\end{cases}
\end{equation}
which takes a piecewise constant form within each half of the driving period. 
By averaging $\epsilon_{K, n}(t)$ over one full cycle, we obtain the average energy $\theta_n(T)$, which provides a unique metric for sorting the Floquet states. 

Numerically, we observe that the Floquet ground state, defined as the state with the lowest average energy $\theta_0(T)$, maintains a high overlap with the ground state of the time-averaged Hamiltonian $(H_0+H_1)/2$ (i.e. the infinite frequency ground state) as long as $T<\pi$. However, this overlap drops sharply to zero once the system enters the heating phase $T>\pi$. This abrupt drop of overlap across the transition point indicates a breakdown of the adiabatic continuity at finite frequency, consistent with the onset of the heating dynamics revealed in our findings.

\section{Effective local quench picture in the non-heating phases}
\label{App:mpscalculations}

As suggested by the average energy spectrum in Fig.~\ref{fig:average_energy}, the non-heating phase at any finite frequency is adiabatically related to the infinite driving frequency cases $T \to 0$, the latter being equivalent to a local quench~\cite{Calabrese_2007}. Thus, in this Appendix, we present numerical results of the entanglement evolution for the gapless NN model obtained from both ED and MPS calculations, which show that the entanglement growth is consistent with the local quantum quench picture. 

\begin{figure}[!htbp]
    \centering
    \includegraphics[width=0.9\linewidth]{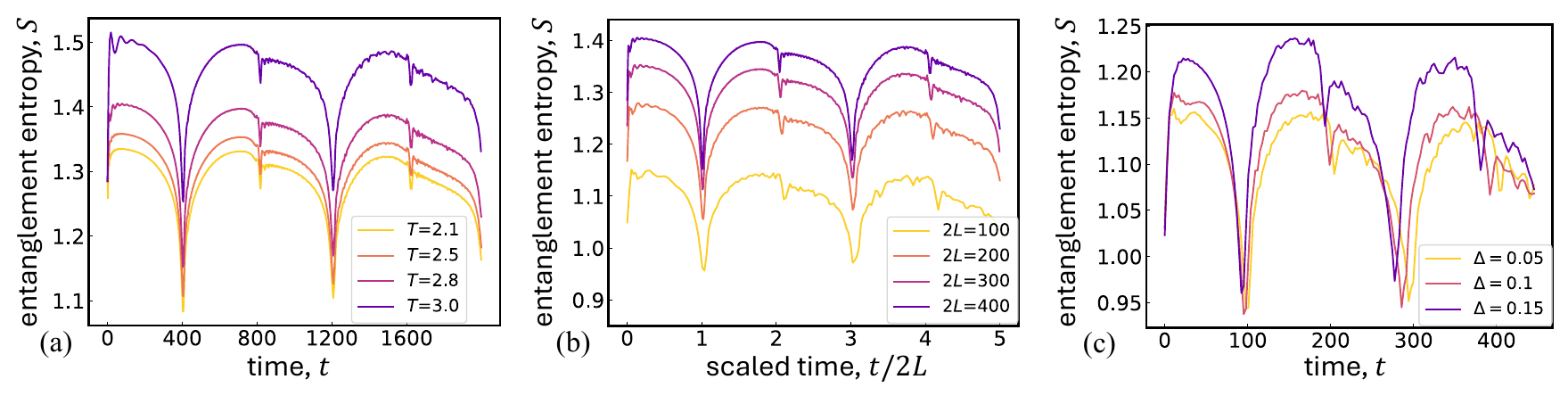}
    \caption{The entanglement evolution in the `local-quench' regime for the gapless NN model under a 2-step driven impurity with $\lambda_0=0.8$. The half-partition entanglement evolution is shown for: the noninteracting limit ($\Delta= 0$), calculated by ED, (a) for various driving periods $T$ with $2L=400$; (b) across different total system sizes $2L$ with $T=2.8$, where the horizontal axis is the scaled time $\tilde{t} = t/(2L)$ and (c) the interacting models, calculated by MPS across different values of the anisotropy $\Delta$ with $T=2.8, 2L=100$. The MPS parameters used are: an initial DMRG bond dimension of $\chi_D = 100$, which determines the ground state and a TEBD bond dimension of $\chi_T=400$ for $\Delta=0.05, 0.1$ and $\chi_T=500$ for $\Delta = 0.15$. }
    \label{fig:revival_evo}
\end{figure}

As shown in Fig.~\ref{fig:revival_evo}(a-b), we find that the half-partition entanglement exhibits partial revivals with a period that is independent of the external driving period $T$, and which scales linearly with the total system size $2L$. Furthermore, the revival period decreases monotonically upon increasing the anisotropy (the NN interaction strength) $\Delta$, as seen in Fig.~\ref{fig:revival_evo}(c).
This entanglement dynamics is similar to that of a local quench, in which case pairs of quasiparticles are propagating from the impurity and reflected at the system's boundary, leading to subextensive growth of entanglement at early times, and revivals at times proportional to the total system size. In fact, the periodicity $\tau$ of the EE takes the form
\begin{equation}
    \tau = \frac{2L}{v}, \quad v = \frac{\pi}{2}\frac{\sqrt{1-\Delta^2}}{\arccos\Delta}, 
    \label{eq:revival_prediction}
\end{equation}
where the velocity of quasiparticles $v$ is known from the Bethe ansatz solution of the (impurity-free) XXZ model~\cite{takahashi1999thermodynamics}. 

\begin{figure}[!htbp]
    \centering
    \includegraphics[width=0.65\linewidth]{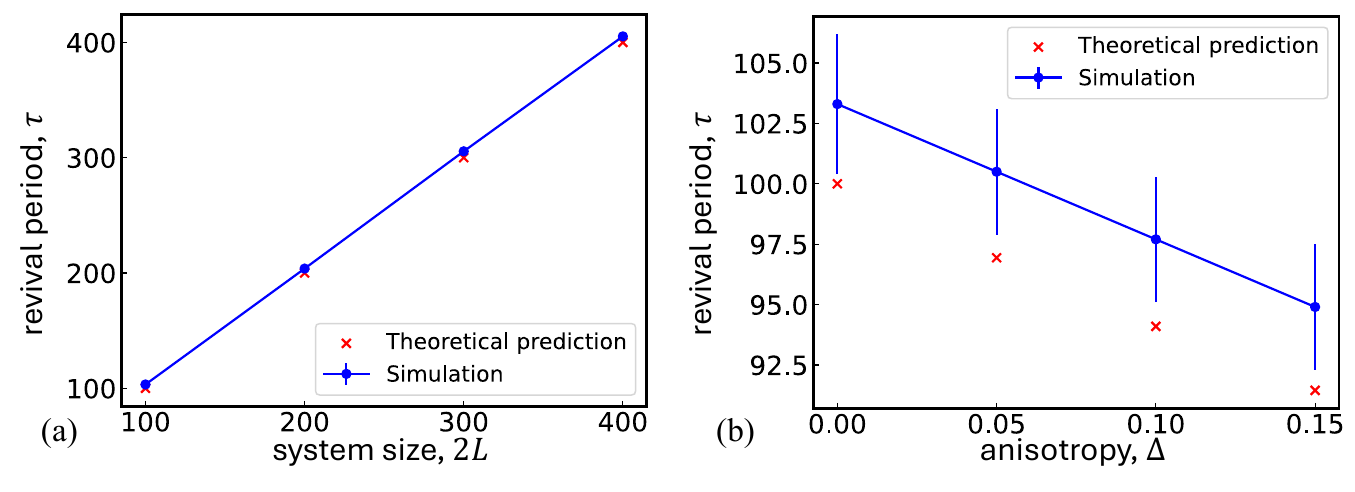}
    \caption{The revival period $\tau$ of the half-partition EE in the non-heating regime for the gapless NN model under a 2-step drive impurity with $T=2.8, \lambda_0=0.8$. (a) The period $\tau$ scales linearly with the total system size $2L$ in the noninteracting limit, and (b) its dependence on the anisotropy $\Delta$ qualitatively follows the trend predicted by the local quench picture with $2L=100$. }
    \label{fig:revival_period}
\end{figure}

We now extract the periodicity of the partial revivals of EE in Fig.~\ref{fig:revival_evo} and compare it with this prediction from the local quantum quench. As demonstrated in Fig.~\ref{fig:revival_period}, $\tau$ showcases linear scaling with the total system size $2L$, closely matching the theoretical prediction of the local quantum quench in the noninteracting limit. Additionally, the dependence of $\tau$ on the anisotropy $\Delta$, can be qualitatively explained by Eq.~\eqref{eq:revival_prediction}, whereas the deviation between the simulation and theoretical prediction may result from finite-size effects. 

\begin{figure}[!htbp]
    \centering
    \includegraphics[width=0.65\linewidth]{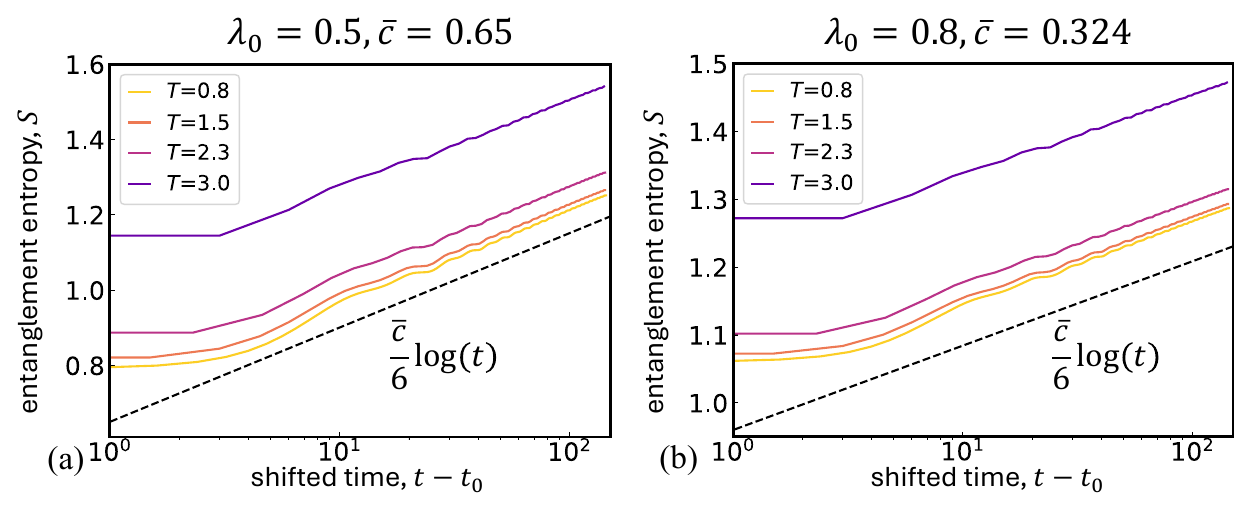}
    \caption{The logarithmic entanglement growth in the non-heating regime for the noninteracting gapless NN model with system size $2L=400$ under a 2-step driven impurity. The half-partition EE is plotted against $\log(t-t_0)$ after the first revival point $t_0 \sim 2L$, exhibiting a linear relationship at late times. Dashed lines represent the averaged slope $\bar{c}/6$ over different driving periods: (a) for $\lambda_0 = 0.5$, $\bar{c} = 0.65 \pm 0.01$ and (b) for $\lambda_0 = 0.8, \bar{c} = 0.324 \pm 0.007$.}
    \label{fig:log_growth}
\end{figure}

Another hallmark of local quench dynamics is the logarithmic growth of EE in time. As shown in Fig.~\ref{fig:revival_evo}(a), the entanglement exhibits a universal logarithmic temporal growth across different driving periods, particularly after the first revival point. 
We investigate this behavior quantitatively in Fig.~\ref{fig:log_growth}, where the EE evolution after the first revival at $t_0 \sim 2L$ is plotted on a logarithmic time axis for various driving periods. A clear linear relationship between EE and $\log(t-t_0)$ emerge at late times. By fitting each curve with the ansatz $c \log (t-t_0) /6 + S_0$, we observe that for a fixed $\lambda_0$, the slope $c$ is approximately independent of the driving period, consistent with the local quench picture.
While $c$ exhibits dependence on the driving strength $\lambda_0$, which can be attributed to a renormalization of the effective central charge by the impurity.

\section{Entanglement transition for non-Hermitian impurity models}
\label{App:nonhermitianfloquet}

In this Appendix, we present the entanglement phases for the gapless NN model with a non-Hermitian 2-step driven impurity as in Eq.~\eqref{eq:floquetdrive_free} with $t_j = 1$ and $\lambda_0> 1$, which corresponds to adding \textit{imaginary} on-site potentials at sites $L$ and $L+1$:
\begin{equation}
H_{\text{imp}}(t) = \frac{1}{2}i\sqrt{\lambda^2(t)-1}[c^{\dagger}_{L}c_L - c^{\dagger}_{L+1}c_{L+1}]-\frac{\lambda(t)}{2}[c^{\dagger}_{L+1}c_L+c^{\dagger}_{L}c_{L+1}].
\label{eq:non-hermitian_impurity}
\end{equation}

\begin{figure}[!htbp]
    \centering
    \includegraphics[width=0.75\linewidth]{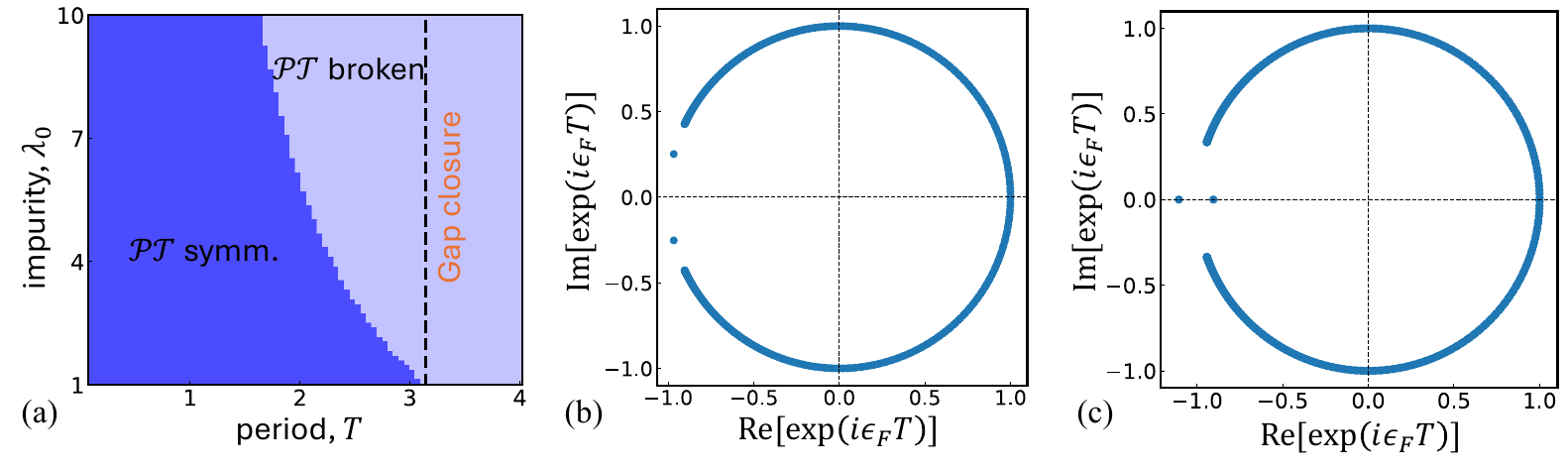}
    \caption{(a) The phase diagram of the gapless NN model under a 2-step driven non-Hermitian impurity with system size $2L=400$ in the $(T, \lambda_0)$ parameter space. The dark blue region corresponds to the PT symmetric phase where the spectrum of the Floquet unitary $U_\text{F}$ lies on a unit circle. The light blue region represents the PT broken phase, where the spectrum of $U_\text{F}$ extends beyond the unit circle. The dashed line marks the bulk spectrum closure point at $T=\pi$, independent of $\lambda_0$. (b)(c) show the spectrum of $U_\text{F}$ for $\lambda_0 = 2$ in the PT symmetric phase at $T=2.7$ and the PT broken phase at $T=2.8$, respectively. }
    \label{fig:PTbreaking}
\end{figure}

For a static impurity with $\lambda(t)\equiv\lambda>1$, the resulting non-Hermitian impurity model showcases parity-time (PT) symmetry~\cite{benderreview} and has a real spectrum which is unitarily related to that of the defect-free chain with $\lambda=1$. 
We now introduce time-dependence via a 2-step drive that alternates periodically, with equal time durations, between the uniform, Hermitian chain ($\lambda(t) = 1$) and the non-Hermitian chain with $\lambda(t)=\lambda_0>1$. 
Although the instantaneous Hamiltonian retains PT symmetry at any time $t$, the resulting Floquet non-Hermitian Hamiltonian could break PT symmetry. 
Such a PT breaking transition is identified through the spectrum of the single-particle Floquet unitary $U_\text{F} = \mathcal{T}\exp(-i\int_0^T \mathrm{d}t H(t))$: in the PT symmetric phase, the eigenvalues $u_n = \exp (i\epsilon_{F, n}T)$ lie on the unit circle, while at the transition they undergo an exceptional point scenario---two complex-conjugate eigenvalues on the circle collide on the real axis and then separate along it, as shown in Fig.~\ref{fig:PTbreaking}. Notably, the PT breaking transition always precedes the closure of the quasienergy bulk gap, with the latter occurring at $T=\pi$; hence, the PT transition point is always located at some $T<\pi$.

\begin{figure}[!htbp]
    \centering
    \includegraphics[width=\linewidth]{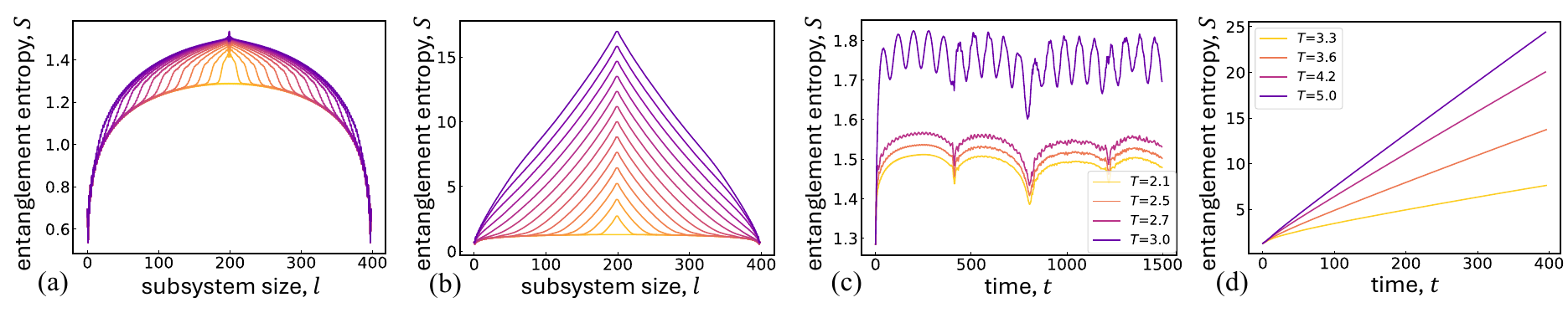}
    \caption{The entanglement phases of the gapless NN model under a 2-step driven non-Hermitian impurity with system size $2L=400$ and $\lambda_0=1.2$. (a)(b) The entanglement scaling (plotted every six Floquet cycles, from orange to purple) in the non-heating phase at $T=2.5$ and the heating phase at $T=4.2$, respectively. (c)(d) The half-partition entanglement evolution across different driving periods $T$ in the non-heating and heating regimes. }
    \label{fig:non-hermitian_evo}
\end{figure}

The entanglement transition, however, remains associated with the closure of the quasienergy bulk gap, as shown in Fig.~\ref{fig:non-hermitian_evo}. As in the Hermitian impurity model, the non-heating phase ($T<\pi$) exhibits logarithmic entanglement scaling over time, with the half-partition EE bounded and showing partial revivals; whereas in the heating phase $(T > \pi)$, the entanglement scales linearly after a transient spreading from the defect and the half partition entanglement showcases a steady linear growth at early times. 

Finally, we note that recent investigations have unveiled purification phase transitions in integrable non-unitary Floquet models with generalized PT-breaking transitions~\cite{zhang2023antiunitarysymmetrybreakinghierarchy}. We therefore seek to determine if similar phenomena manifest in our model.
In fact, in the PT symmetric phase, the Floquet spectrum sits on the unit circle, precluding the dominance of any single Floquet eigenmode, such that any initially mixed state persists to be mixed. 
On the other hand, in the PT broken phase, if there exists a single eigenvalue in the many-body Floquet operator that has the largest absolute value, it will drive the purification of any initial mixed state towards the associated Floquet eigenstate. 
However, our findings in the non-Hermitian driven impurity model deviate from this empirical behavior.
Instead, we observe that any initial mixed state will remain \textit{mixed} even within the PT-broken regime, because there are always exponentially many Floquet eigenstates with the same largest absolute value in the many-body Floquet operator. Specifically, at half-filling, there are approximately $4^{L}/\sqrt{2L}$ of such modes. 
Consequently, to induce purification phase transitions in such systems, it would be advantageous to incorporate interaction terms that preserve the PT symmetry at the level of the Floquet Hamiltonian. These interactions would serve to lift the degeneracy of the Floquet eigenvalues and lead to purification phase transitions.

\section{Entanglement transition for various impurity configurations}
\label{App:universality}

In this Appendix, we demonstrate that the entanglement transition reported in the main text is not specific to the conformal defect, but persists across qualitatively different impurity configurations. To this end, we consider three alternative forms of the driven impurity configuration $\Lambda(t)$ in the noninteracting gapless NN chain. The total Hamiltonian takes the form of Eq.~\eqref{eq:floquetdrive_free} with $t_j = 1$, 
\begin{equation}
    H(t) = - \frac{1}{2} \sum_{j=1}^{L-1} (c^\dagger_j c_{j+1} + h.c.) - \frac{1}{2} \sum_{j=L+1}^{2L-1} (c^\dagger_j c_{j+1} + h.c.) + \left(c^{\dagger}_L,\ c^{\dagger}_{L+1} \right)\Lambda(t)\begin{pmatrix} c_L \\ c_{L+1} \end{pmatrix},
\end{equation}
and subject to a 2-step drive with $\lambda_0 = 0.5$. In each case, we initialize the system from the half-filling ground state of the impurity-free chain and monitor the half-partition EE.

\begin{figure}[!htbp]
    \centering
    \includegraphics[width=0.7\linewidth]{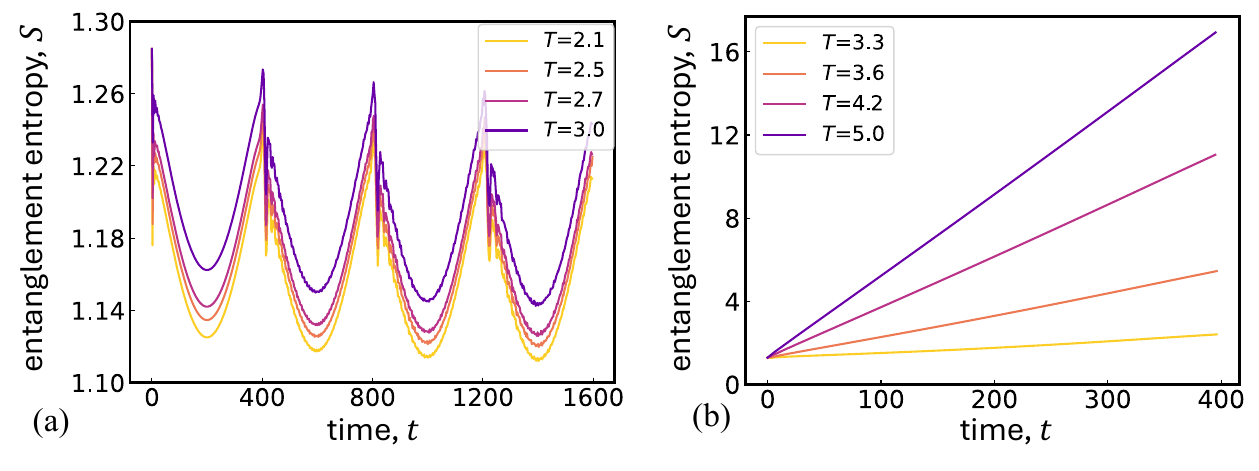}
    \caption{The half-partition EE evolution of the gapless NN chain with $2L=400$, subject to a hopping impurity with $\lambda_0=0.5$ for (a) the non-heating regime ($T < T_\ast$), and (b) the heating regime ($T>T_\ast$).}
    \label{fig:hopping_evo}
\end{figure}

\paragraph{Hopping impurity.} The first configuration modulates only the hopping amplitude across the central bond, with
\begin{equation}
    \Lambda(t) = - \frac{1}{2} 
    \left(
    \begin{array}{cc}
       0  &  \lambda(t) \\
       \lambda(t)  &  0
    \end{array}
    \right),
\end{equation}
which corresponds to a driven quantum point contact. Unlike the conformal defect, this impurity does not introduce any on-site potential. As shown in Fig.~\ref{fig:hopping_evo}, the entanglement transition at $T_\ast = \pi$ persists: the EE remains bounded with periodical partial revivals for $T < T_\ast$ while grows linearly for $T>T_\ast$. 

\begin{figure}[!htbp]
    \centering
    \includegraphics[width=0.7\linewidth]{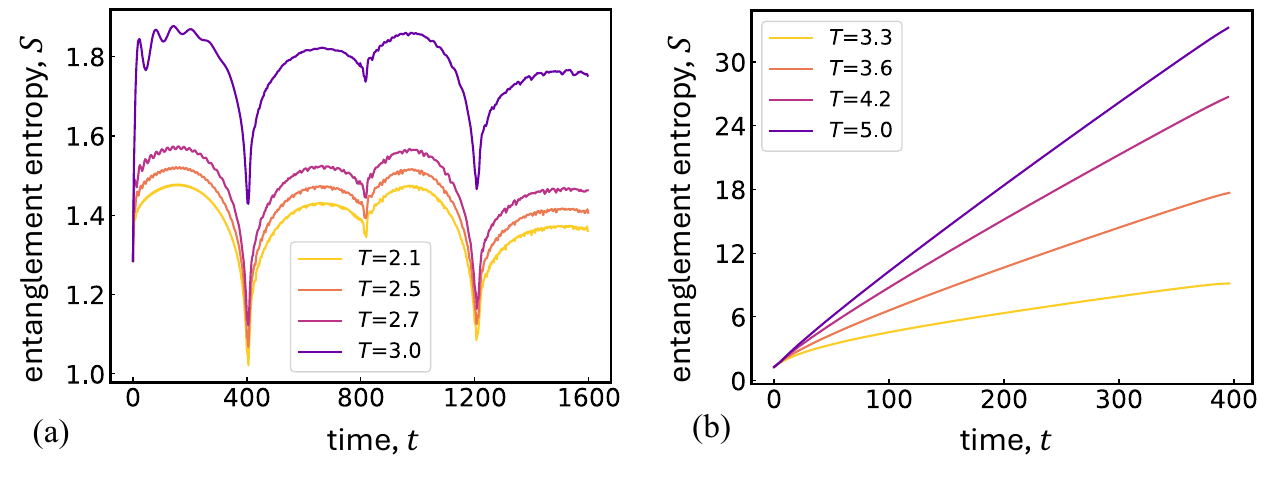}
    \caption{The half-partition EE evolution of the gapless NN chain with $2L=400$, subject to a double impurity with $\lambda_0=0.5$ for (a) the non-heating regime ($T < T_\ast$), and (b) the heating regime ($T>T_\ast$).}
    \label{fig:double_evo}
\end{figure}

\paragraph{Double impurity.} This configuration drives the onsite energies of sites $L$ and $L+1$ while keeping the hopping amplitude static,
\begin{equation}
    \Lambda(t) = \frac{1}{2} 
    \left(
    \begin{array}{cc}
       \sqrt{1-\lambda^2(t)}  &  -1 \\
       -1  &  -\sqrt{1-\lambda^2(t)}
    \end{array}
    \right) ,
\end{equation}
Here, the off-diagonal elements are identical to the bulk hoppings and the drive enters only through the diagonal (onsite) terms. As shown in Fig.~\ref{fig:double_evo}, the same transition at $T_\ast = \pi$ is observed, confirming that the mechanism does not rely on modulating the inter-site coupling.

\paragraph{Single impurity. } Finally, we consider the case where the drive acts on only one of the two central sites,
\begin{equation}
    \Lambda(t) = \frac{1}{2} 
    \left(
    \begin{array}{cc}
       \sqrt{1-\lambda^2(t)}  &  -1 \\
       -1  &  0
    \end{array}
    \right) 
\end{equation}
which differs from the double impurity by the absence of the time-dependent on-site potential on site $L+1$. Fig.~\ref{fig:sin_evo} again confirms the robustness of the transition at $T_\ast = \pi$, suggesting that the phenomenon relies neither on the mirror symmetry of the impurity nor on the number of driven sites, as long as the drive remains local.

\begin{figure}[!htbp]
    \centering
    \includegraphics[width=0.7\linewidth]{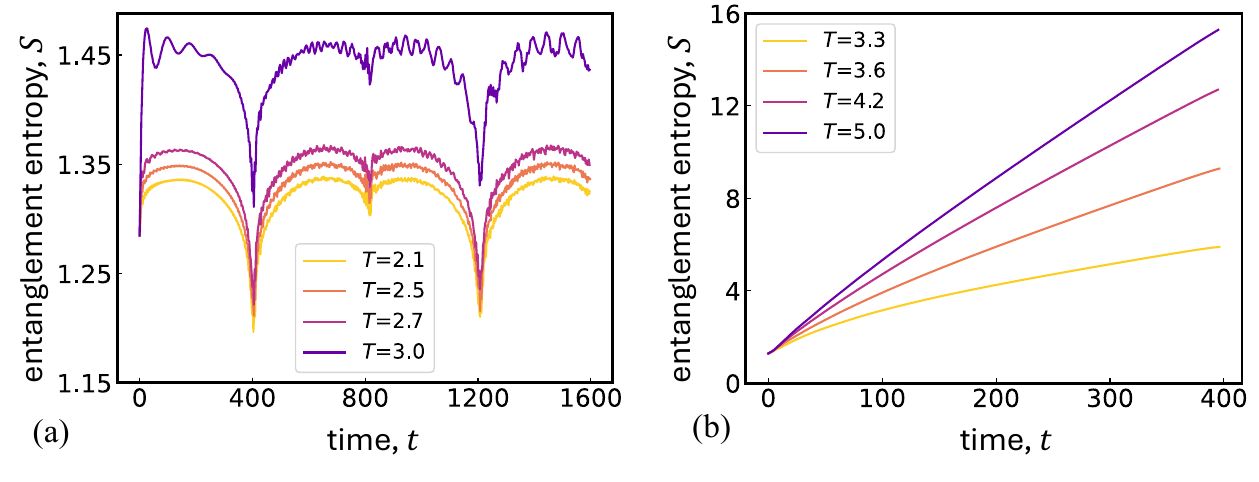}
    \caption{The half-partition EE evolution of the gapless NN chain with $2L=400$, subject to a single impurity with $\lambda_0=0.5$ for (a) the non-heating regime ($T < T_\ast$), and (b) the heating regime ($T>T_\ast$).}
    \label{fig:single_evo}
\end{figure}

Taken together, these results demonstrate that the entanglement transition at $T_\ast = \pi$ is a universal feature of locally driven free-fermion chains: it depends only on the bulk bandwidth and the driving frequency, and is insensitive to the microscopic details of the impurity configuration. 

\section{Convergence and Benchmarks of MPS Simulations}
\label{APP:MPS_backup}

This Appendix presents numerical benchmarks and convergence checks supporting the reliability of the MPS simulations presented in Fig.~\ref{fig:MPS_evo}. 
Our MPS simulation pipeline consists of two stages: first, we employ DMRG to find the half-filling many-body ground state of the interacting model, which serves as our initial state; second, we simulate the Floquet dynamics by propagating this state using TEBD. 

\begin{figure}[!htbp]
    \centering
    \includegraphics[width=0.9\linewidth]{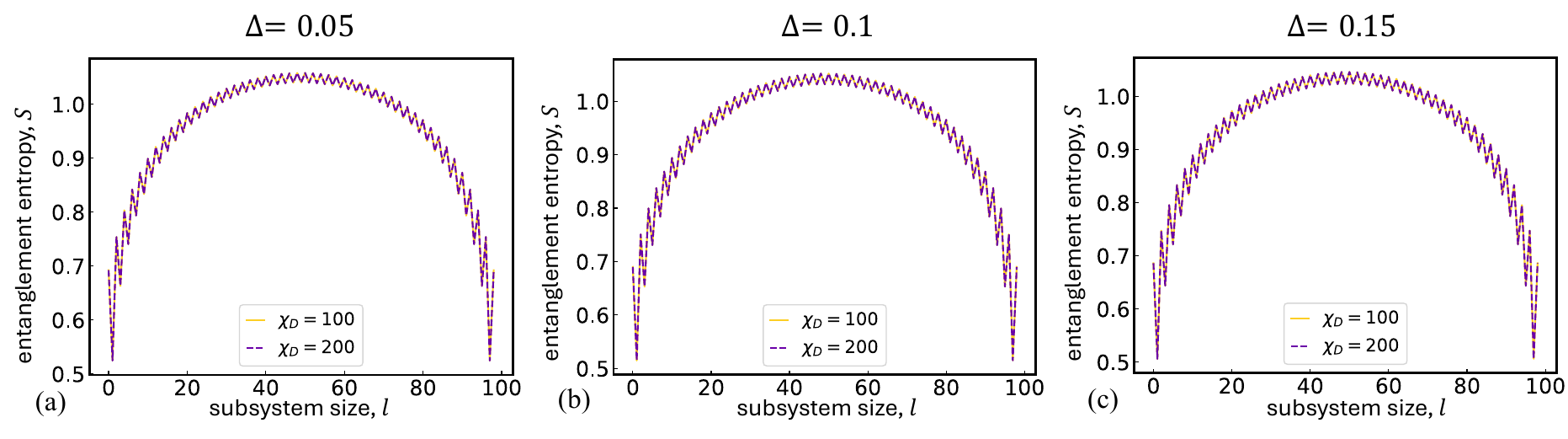}
    \caption{Convergence of initial state entanglement profiles with respect to the DMRG bond dimension $\chi_D$. The initial states are prepared from the interacting gapless model with $2L=100$ using DMRG across different interaction strengths (a) $\Delta=0.05$, (b) $\Delta = 0.1$ and (c) $\Delta=0.15$. By comparing the spatial entanglement scaling across different values of $\chi_D$, the plots illustrate that the DMRG algorithm achieved strong convergence by $\chi_D \sim 100$.}
    \label{fig:dmrg_convergence}
\end{figure}

Following this streamline, we first verify that the bond dimension limit for the DMRG step is sufficient to provide a well-converged initial state. To establish this, we compute the entanglement scaling of the initial state under varying bond dimension cutoffs $\chi_D$ across all interacting strengths $\Delta$ presented in Fig.~\ref{fig:MPS_evo}. As shown in Fig.~\ref{fig:dmrg_convergence}, the ground-state entanglement structure exhibits excellent convergence with respect to $\chi_D$, thus we conclude that setting $\chi_D=100$ is sufficient to achieve high numerical accuracy for the initial state.

\begin{figure}[!htbp]
    \centering
    \includegraphics[width=0.9\linewidth]{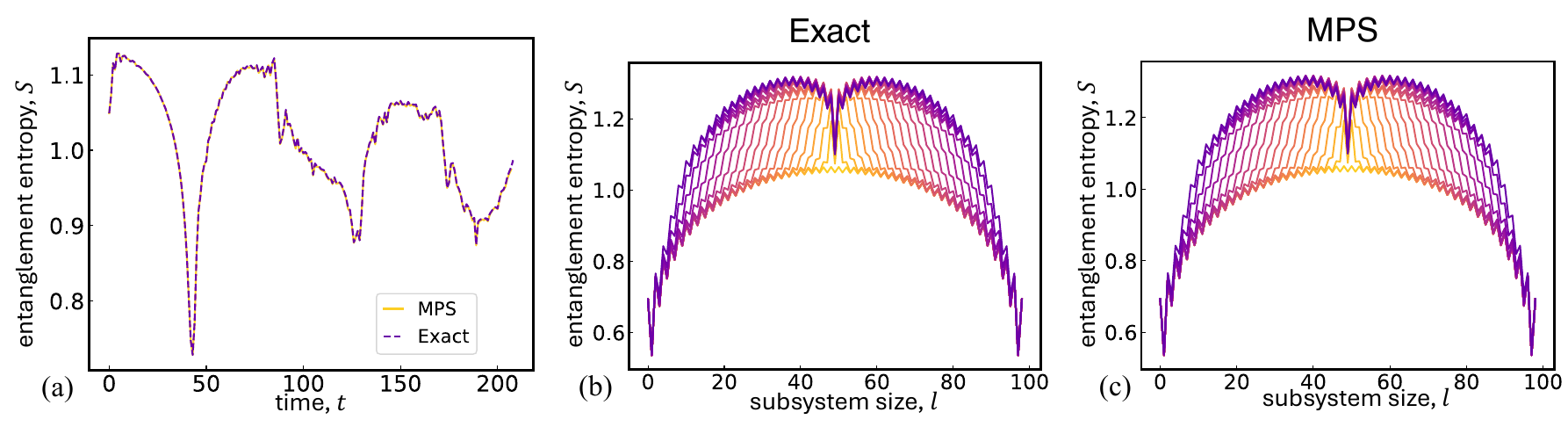}
    \caption{Benchmarks on the TEBD algorithm. The entanglement dynamics are simulated in the noninteracting gapless model with $2L=100$ under a 2-step drive at $T=2.4$ and $\lambda_0=0.5$ using both exact correlation-matrix method and the MPS method. For the MPS simulations, the initial state is prepared via DMRG with $\chi_D=100$. (a) The comparison of half-partition EE evolution across both methods. The entanglement scalings (plotted every cycle, from yellow to purple) at different times are computed via (b) the exact correlation-matrix method and (c) the MPS method.}
    \label{fig:TEBD_bench_noninteracting}
\end{figure}

Next, to ensure that our TEBD protocol faithfully implements the targeted Floquet evolution, we benchmark our TEBD script against exact, noninteracting ($\Delta=0$) results obtained via the free-fermion correlation-matrix method. This comparison is shown in Fig.~\ref{fig:TEBD_bench_noninteracting}, where the entanglement profiles calculated via both methods are in good agreement, confirming the accuracy of our time-evolution script.

\begin{figure}[!htbp]
    \centering
    \includegraphics[width=0.9\linewidth]{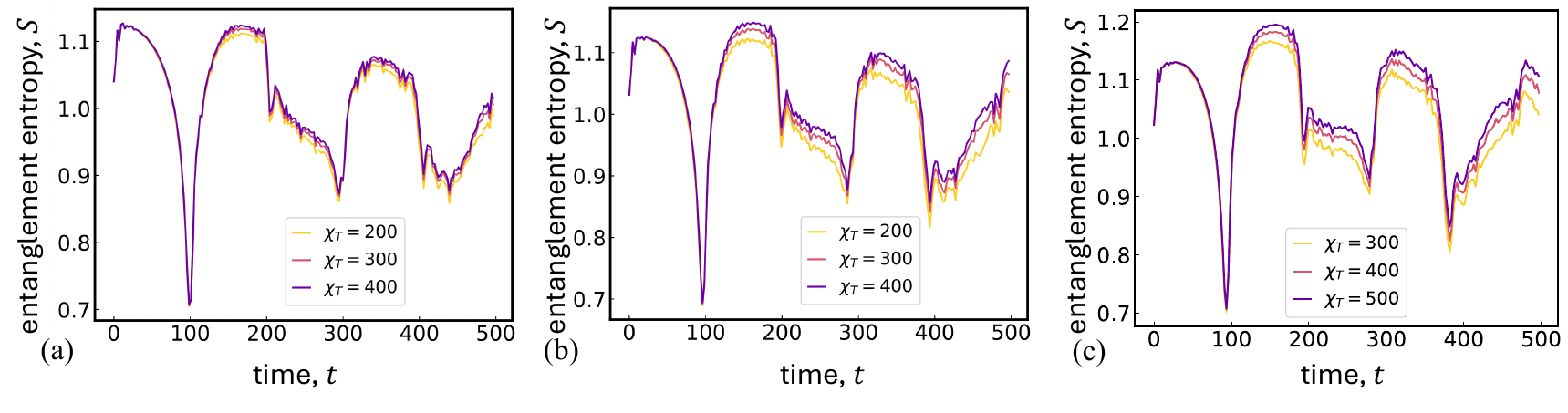}
    \caption{Convergence of entanglement dynamics with respect to the TEBD bond dimension $\chi_T$. The half-partition entanglement evolution is simulated in the interacting gapless model with $2L=100$, under a 2-step drive at $T=2.4$ and $\lambda_0=0.5$. The initial states are prepared via DMRG with $\chi_D = 100$. Results are plotted across several values of $\chi_T$ for interaction strengths (a) $\Delta=0.05$, (b) $\Delta=0.1$, and (c) $\Delta = 0.15$. 
    The close overlap of curves validates the bond dimension truncation used in the main text. }
    \label{fig:TEBD_XXZ}
\end{figure}

Finally, we systematically test the convergence of the entanglement dynamics within the interacting regime ($\Delta>0$) with respect to the TEBD bond dimension limit $\chi_T$. To achieve this, we simulate the half-partition entanglement dynamics in the interacting model ($\Delta > 0$) under several values of $\chi_T$, and inspect their differences. As presented in Fig.~\ref{fig:TEBD_XXZ}, the finite-time dynamics display convergence beyond a critical threshold of $\chi_T$. This comparison further justifies our parameter choice of $\chi_T$ in Fig.~\ref{fig:MPS_evo}, ensuring that our physical conclusions are numerically stable. 

\section{Floquet Hamiltonian of the driven impurity model}
\label{App:floquetham}

In this Appendix, we demonstrate how to analytically derive the Floquet Hamiltonian for the gapless NN model with a harmonically driven impurity using the rotating frame method.
The time-dependent Hamiltonian, as introduced in Eq.~\eqref{eq:floquetdrive_free} consists of two decoupled homogeneous noninteracting fermion chains of length $L$, connected by a two-site impurity subject to periodic driving: 
\begin{equation}
    H(t) = H_L+H_R + H_{\text{imp}}(t),
\end{equation}
with
\begin{equation}
    H_L = - \frac{1}{2}\sum_{j=1}^{L-1} (c^\dagger_j c_{j+1}+h.c.), \quad H_R = -\frac{1}{2} \sum_{j=L+1}^{2L-1} (c^\dagger_j c_{j+1}+h.c.)
\end{equation}
and the driven impurity is given by
\begin{equation}
     H_{\text{imp}}(t) = \frac{1}{2} (c^\dagger_L, c^\dagger_{L+1}) \left(
    \begin{array}{cc}
        \sin(2\pi t/T)  &  -\cos (2\pi t/T) \\
      -\cos (2\pi t/T)    &  -\sin(2\pi t/T)
    \end{array}\right) \left(
    \begin{array}{c}
        c_L   \\
        c_{L+1}
    \end{array}\right).
\end{equation}
The idea of the rotating frame method is to design a time-periodic micromotion operator $P(t)$ such that the effective Hamiltonian $\tilde{H}$ becomes a time-independent one,
\begin{equation}
    \tilde{H} = P^\dagger(t) H_t P(t) - i P^\dagger(t)  \frac{\partial P(t)}{\partial t} .
\end{equation}
To explicitly construct $P(t)$, we exploit the following algebraic relations:
\begin{equation}
    [\sigma, H_L+H_R] = 0, \quad [\sigma, \Omega] = -2i \Gamma, \quad [\sigma, \Gamma] = 2i \Omega
\end{equation}
where $\sigma=i\sum_{j=1}^{L}[c^{\dagger}_jc_{2L+1-j}-h.c.]$ is the mirror operator, and $\Omega = c^\dagger_L c_L - c^\dagger_{L+1} c_{L+1}, \Gamma = c^\dagger_L c_{L+1} + c^\dagger_{L+1} c_L $. Crucially, we note that the operators $\{\sigma,\Gamma,\Omega\}$ form an SU(2) algebra.
Using this algebraic structure, we can explicitly design the micromotion operator, $P(t) = \exp (i\pi t/ T (\sigma - N))$, where $N$ denotes the total particle number, which yields the following effective time-independent Hamiltonian, 
\begin{equation}
    \tilde{H} = H(0) + \frac{\pi}{T}(\sigma-N),
\end{equation}
where $H(0)$ is the Hamiltonian of a homogeneous noninteracting fermion chain of length $2L$. 
Noting that $\exp(i\pi \sigma) = \exp(i\pi N)$ is the parity operator, the micromotion operator $P(t)$ also satisfies the time periodicity.
Thereby, the corresponding Floquet Hamiltonian is given exactly by: 
\begin{equation}
    H_{\text{F}} = P(0) \tilde{H} P(0)^\dagger = H(0) + \frac{\pi}{T} (\sigma - N). 
    \label{eq:floquet_ham}
\end{equation}
We stress that this result is exact and non-perturbative. This tells us that the Floquet Hamiltonian consists of the nearest neighbor hoppings as well as non-local rainbow-like long-range hoppings between sites $i$ and $2L+1-i$. This long-range part of the Floquet Hamiltonian is responsible for the entanglement transition at $T=\pi$. 

\begin{figure}[!htbp]
    \centering
    \includegraphics[width=\linewidth]{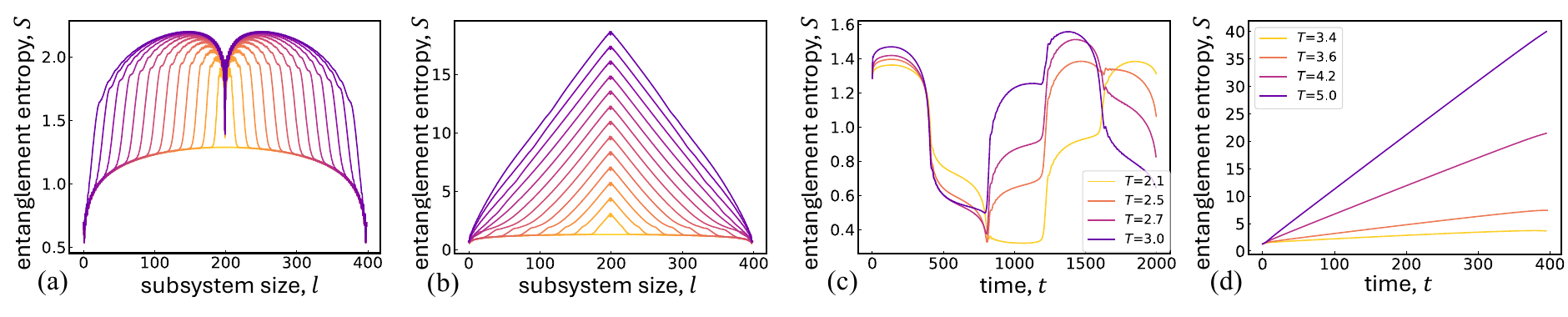}
    \caption{The entanglement phases for the harmonically driven impurity model with system size $2L=400$. (a)(b) The entanglement scaling (plotted every six Floquet cycles, from orange to purple) in the non-heating phase at $T=2.5$ and the heating phase at $T=4.2$, respectively. (c)(d) The half-partition entanglement evolution across different driving periods $T$ in the non-heating and heating regimes. 
     }
    \label{fig:sin_evo}
\end{figure}

Fig.~\ref{fig:sin_evo} presents the entanglement evolution of the harmonically driven impurity model, which bears similar behavior to that of the 2-step driven impurity model.
In the heating phase $(T > \pi)$, the entanglement follows a linear scaling after an initial transient spreading from the defect. Correspondingly, the half partition entanglement shows a steady linear growth at early times. 
On the other hand, in the non-heating phase ($T<\pi$), the EE exhibits logarithmic scaling over time and the half-partition EE remains bounded. Differing from the 2-step driven case, the revival period of the half-partition entanglement here depends on the driving frequency.

In addition, the single particle spectrum and eigenstates of the Floquet Hamiltonian Eq.~\eqref{eq:floquet_ham} can be analytically derived~\cite{PhysRevB.103.L041405}.  
To do so, we first start by writing down the eigen-equations in both left and right chains $(1\le j < L)$:   
\begin{equation}
    E\psi_j =- \frac{1}{2}\psi_{j-1} - \frac{1}{2}\psi_{j+1} +i \frac{\pi}{T}\psi_{2L+1-j} - \frac{\pi}{T} \psi_j, \quad E\psi_{2L+1-j} = -\frac{1}{2}\psi_{2L-j} -\frac{1}{2}\psi_{2L+2-j} -i\frac{\pi}{T}\psi_{j}  - \frac{\pi}{T} \psi_{2L+1-j},
\end{equation}
which can be easily solved by considering the following even/odd combinations: $\phi_j^\pm = \psi_j \mp i \psi_{2L+1-j}$. 
Then the eigen-equations can be rewritten in terms of $\phi_j^{\pm}$ as:
\begin{equation}
    (2E + \frac{2\pi}{T}  \pm \frac{2\pi}{T})  \phi_j^{\pm} + \phi_{j+1}^\pm + \phi_{j-1}^\pm = 0
\end{equation}
whose general solutions are exponential series,
\begin{equation}
    \phi_j^\pm = A_\pm e^{\kappa_\pm j} + B_\pm e^{-\kappa_\pm j}, \quad -\cosh \kappa_\pm = E + \frac{\pi}{T}\pm \frac{\pi}{T}.
    \label{eq: general_wavefunction}
\end{equation}
The coefficients can be fixed by matching the boundary conditions, 
\begin{equation}\left\{
    \begin{aligned}
        & \phi_0^\pm = 0 \\
        & (2E + \frac{2\pi}{T}) \phi_L^+ = - \phi_{L-1}^+ + i \phi_L^- - \frac{2\pi}{T} \phi_L^+  \\
        & (2E + \frac{2\pi}{T}) \phi_L^- = - \phi_{L-1}^- - i \phi_L^+ + \frac{2\pi}{T} \phi_L^- \\
    \end{aligned}\right.
\end{equation}
by plugging in the expressions of Eq.~\eqref{eq: general_wavefunction}, we derive that the eigenvalues of the Floquet Hamiltonian Eq.~\eqref{eq:floquet_ham} is determined by the solutions of the following equation,
\begin{equation}
    \frac{\sinh (\kappa_+ L)}{\sinh (\kappa_+ (L+1))} - \frac{\sinh(\kappa_-(L+1))}{\sinh(\kappa_- L)} = 0
\end{equation}
with the corresponding eigenvector provided as,
\begin{equation}
\begin{aligned}
    &|E\rangle = \frac{1}{\sqrt{N(E)}}\sum_{j=1}^L (\frac{\sinh(\kappa_+ j)}{\sinh(\kappa_+(L+1))} + i \frac{\sinh(\kappa_- j)}{\sinh(\kappa_- L)})  c^\dagger_j |\mathrm{vac}\rangle \\
    &+ \frac{i}{\sqrt{N(E)}} \sum_{j=L+1}^{2L} (\frac{\sinh(\kappa_+ (2L+1-j))}{\sinh(\kappa_+(L+1))} - i \frac{\sinh(\kappa_- (2L+1-j))}{\sinh(\kappa_- L)}) c^\dagger_j |\mathrm{vac}\rangle
\end{aligned}
\label{eq:Floquet_eigenbasis}
\end{equation}
where $N(E)$ is the normalization factor. 


\section{Schrieffer-Wolff transformation of the Floquet Hamiltonian}
\label{App:schriefferwolff}

Here, we apply the Schrieffer-Wolff (SW) transformation to derive an effective low-energy Hamiltonian for the driven dynamics in the high-frequency limit $(T \to 0)$. 
We start from the Floquet Hamiltonian in Eq.~\eqref{eq:floquet_ham}. 
In the high-frequency regime, the nearest neighbor hopping terms in $H(0)$ can be treated perturbatively. 
To proceed, we first note that the mirror operator $\sigma$ is diagonalized as
\begin{equation}
\sigma= \sum_{j=1}^L [ \gamma_{j,+}^{\dagger}\gamma_{j,+}-\gamma_{j,-}^{\dagger}\gamma_{j,-}  ],
\end{equation}
where we introduced the (anti)-bonding operators $\gamma_{j,\pm}=\frac{1}{\sqrt{2}}[c_j\pm i c_{2L+1-j}]$. From here, it is clear that the spectrum of $\sigma$ consists of two flat bands, with eigenvalues $\pm1$.
The next step is to express the uniform hopping Hamiltonian $H(0)$ in terms of the bonding operators.  It reads
\begin{equation}
H(0) = - \frac{1}{2}\left(\sum_{j=1, \sigma=\pm}^{L-1}[\gamma_{j+1,\sigma}^{\dagger}\gamma_{j,\sigma}+\gamma_{j,\sigma}^{\dagger}\gamma_{j+1,\sigma}]\right) + \frac{i}{2}[\gamma_{L,-}^{\dagger}\gamma_{L,+}-\gamma_{L,+}^{\dagger}\gamma_{L,-}].
\end{equation}
Therefore, we note that in the bulk, the hopping Hamiltonian does not mix the different chirality sectors, which is in agreement with the commutator $[H_L+H_R,\sigma]=0$, whereas the defect term at $j=L$ does mix the two sectors. 
Our goal is to project the Hamiltonian on the lower energy space with projector $P_- = \sum_{j=1}^L\gamma_{j,-}^{\dagger}|0\rangle \langle 0|\gamma_{j,-}$. 
To this end, we split the perturbation term as $H(0)=H_{\text{diag}}+H_{\text{non-diag}}$, where the second term mixes the two chiralities, while the first term does not. 
We need to perturbatively find an operator $S$ such that the resultant Hamiltonian $\tilde{H} = e^S H_\text{F} e^{-S}$ is block diagonal in terms of bonding operators. 
To this end, we expand $S$ in series of $T$:
\begin{equation}
    S = TS^{(0)} + T^2 S^{(1)} + T^3 S^{(2)}+ \mathcal{O}(T^4)
    \label{eq:S_expand}
\end{equation}
Therefore, we can obtain the effective Hamiltonian by expanding order by order as: 
\begin{equation}
\begin{gathered}
\tilde{H}= e^{S}H_{\text{F}}e^{-S}=\frac{\pi}{T}(\sigma-N)  + (H_{\text{diag}} + H_\text{non-diag} + \pi [S^{(0)}, \sigma]) \\
+ (T [S^{(0)},H_{\text{diag}} + H_\text{non-diag}]+\pi T[S^{(1)},\sigma]+ \frac{\pi T}{2} [S^{(0)}, [S^{(0)}, \sigma]] )\\
+(\frac{\pi T^2}{6} [S^{(0)}, [S^{0}, [S^{(0)}, \sigma]]] + \frac{T^2}{2} [S^{(0)}, [S^{(0)}, H_{\text{diag}} + H_\text{non-diag}]] + T^2 [S^{(1)}, H_{\text{diag}} + H_\text{non-diag}] \\
+ \pi T^2 [S^{(2)}, \sigma] + \frac{\pi T^2}{2}[S^{(0)}, [S^{(1)}, \sigma]]+\frac{\pi T^2}{2}[S^{(1)}, [S^{(0)}, \sigma]]) + \mathcal{O}(T^3),
\end{gathered}
\label{eq:SW_expand}
\end{equation}
To cancel the non-diagonal terms at the leading order, we thereby require,
\begin{equation}
\pi [S^{(0)}, \sigma]=-H_{\text{non-diag}} =-\frac{i}{2} [\gamma_{L,-}^{\dagger}\gamma_{L,+}-\gamma_{L,+}^{\dagger}\gamma_{L,-}].
\end{equation}
Exploiting the algebraic structure noted in the last section, we thus obtain
\begin{equation}
S^{(0)}= -\frac{i}{4 \pi}[\gamma_{L,-}^{\dagger}\gamma_{L,+}+\gamma_{L,+}^{\dagger}\gamma_{L,-}]. 
\end{equation}
Then Eq.~\eqref{eq:SW_expand} can be simplified as,
\begin{equation}
\begin{gathered}
\tilde{H}= e^{S}H_{\text{F}}e^{-S}=\frac{\pi}{T}(\sigma-N)  + H_{\text{diag}}  + \frac{T}{8\pi}[\gamma_{L,+}^{\dagger}\gamma_{L,+}-\gamma_{L,-}^{\dagger}\gamma_{L,-}]\\
+(\frac{\pi T^2}{3} [S^{(0)}, [S^{0}, H_\text{non-diag}]] + \frac{T^2}{2} [S^{(0)}, [S^{(0)}, H_{\text{diag}}]] + T^2 [S^{(1)}, H_{\text{diag}} + \frac{1}{2} H_\text{non-diag}] \\
+ \pi T^2 [S^{(2)}, \sigma] + \frac{\pi T^2}{2}[S^{(0)}, [S^{(1)}, \sigma]])+ \mathcal{O}(T^3), \\
\end{gathered}
\label{eq:SW_first_order}
\end{equation}
where we use the following commutation relations: 
\begin{equation}
\begin{aligned}
    &[S^{(0)},H_{\text{non-diag}}]= \frac{1}{4\pi}[\gamma_{L,+}^{\dagger}\gamma_{L,+}-\gamma_{L,-}^{\dagger}\gamma_{L,-}], \\
    &[S^{(0)}, H_{\text{diag}}] = \frac{i}{8\pi} (\gamma_{L, +}^\dagger \gamma_{L-1, -} - \gamma^\dagger_{L-1, +} \gamma_{L, -}- h.c.),
\end{aligned}
\end{equation}
and the second one yields a purely block off-diagonal term, which will be eliminated by the commutator $\pi [S^{(1)}, \sigma]$, which further gives,
\begin{equation}
    S^{(1)} = \frac{i}{16 \pi^2} (\gamma_{L, +}^\dagger \gamma_{L-1, -} - \gamma^\dagger_{L-1, +} \gamma_{L, -} + h.c.)
\end{equation}
Following a similar procedure, we can derive the effective Hamiltonian up to the order of $T^2$, 
\begin{equation}
\tilde{H}=\frac{\pi}{T}(\sigma-N)  + H_{\text{diag}}  + \frac{T}{8\pi}[\gamma_{L,+}^{\dagger}\gamma_{L,+}-\gamma_{L,-}^{\dagger}\gamma_{L,-}] + \frac{T^2 }{64 \pi^2} (\gamma^\dagger_{L, -} \gamma_{L-1, -} + \gamma^\dagger_{L, +} \gamma_{L-1, +} + h.c.) + \mathcal{O}(T^3), 
\label{eq:SW_simplified}
\end{equation}
Then, by projecting to the lower energy sector, the effective Hamiltonian is eventually given as,
\begin{equation}
H_{\text{eff}}=P_-\tilde{H}  P_-= -\frac{2\pi}{T}\sum_{j=1}^L \gamma_{j,-}^{\dagger}\gamma_{j,-}- \frac{1}{2}\sum_{j=1}^{L-1} [\gamma^{\dagger}_{j+1,-}\gamma_{j,-}+\gamma^{\dagger}_{j,-}\gamma_{j+1,-}]-\frac{T}{8\pi}\gamma_{L,-}^{\dagger}\gamma_{L,-} + \frac{T^2 }{64 \pi^2} (\gamma^\dagger_{L, -} \gamma_{L-1, -} + h.c.) 
\label{eq:SW_H_eff}
\end{equation}

Based on Eq.~\eqref{eq:SW_H_eff}, we can understand the role of successive perturbative orders:
\begin{enumerate}
\setcounter{enumi}{-1}
    \item At zeroth order, the Floquet Hamiltonian reduces $\pi (\sigma - N)/T$, such that the stroboscopic dynamics is given by $e^{-i \pi n (\sigma-N)}=1$, thus EE remains constant over time, $S_A(t)=S_A(0)$.
\item At first order, only bulk hopping terms are generated, with no coupling at the defect. This implies that the dynamics is effectively governed by $H_L + H_R$, which corresponds to two uncoupled chains. 
This scenario is basically a local quantum quench, where the entanglement will only oscillate as function of time (in particular, the half-system entanglement remains constant in this case).
\item At second order, a local on-site perturbation appears at the defect site. We can expect that this purely local term should not have significant effect on the entanglement dynamics 
\item At third order, the perturbation is still localized around the defect, but now couples sites $L-1$ and $L+1$, as well as $L$ and $L+2$. These higher-order terms gradually extend the range of defect-induced couplings, though the growth is slow with increasing order.
\end{enumerate}

In conclusions, the SW transformation demonstrates that, in the high-frequency limit, the effective dynamics is governed primarily by local perturbations around the defect. The spatial range of these perturbations increases only gradually with perturbative orders, indicating that the defect’s influence spreads slowly into the bulk.

\end{document}